\documentclass{pas}

\usepackage{multirow}
\usepackage{amssymb}
\usepackage{lineno}

\begin{document}
% \linenumbers

\lefttitle{Publications of the Astronomical Society of Australia}
\righttitle{Gannon, Ferr\'e-Mateu and Forbes}

\jnlPage{1}{4}
\jnlDoiYr{2025}
\doival{10.1017/pasa.xxxx.xx}

\articletitt{The Dawes Review}

\title{The Dawes Review 14: A Decade of Ultra-Diffuse Galaxies}

\author{\gn{Jonah S.} \sn{Gannon}$^1$, \gn{Anna} \sn{Ferr\'e-Mateu},$^{2,3,1}$ and \gn{Duncan A.} \sn{Forbes}$^{1}$}

\affil{$^{1}$ Centre for Astrophysics \& Supercomputing, Swinburne University, Hawthorn, VIC 3122, Australia\\
$^{2}$ Instituto de Astrof\'isica de Canarias, Av. Via Lactea s/n, E38205 La Laguna, Spain\\
$^{3}$ Departamento de Astrof\'isica, Universidad de La Laguna, E-38200, La Laguna, Tenerife, Spain}

\corresp{J.S. Gannon, Email: jonah.gannon@gmail.com}

\citeauth{Gannon J.S, Ferr\'e-Mateu A. and Forbes D.A., The Dawes Review 14: A Decade of Ultra-Diffuse Galaxies {\it Publications of the Astronomical Society of Australia} {\bf 00}, 1--12. https://doi.org/10.1017/pasa.xxxx.xx}

\history{(Received xx xx xxxx; revised xx xx xxxx; accepted xx xx xxxx)}

\begin{abstract}
It has been 10 years since the initial discovery of ``Ultra-Diffuse Galaxies'' (UDGs) in the Coma cluster and the revelation that {large}, low surface brightness galaxies may constitute a {greater} fraction of galaxies than first thought. This left an open question: Are UDGs something special, or just an extension of the previously known dwarf galaxy population? Seeking to answer this question, in the decade following, dedicated simulations have studied and proposed a myriad of formation pathways to create UDGs. Observations have then pushed the limits of world-class observatories to perform detailed studies of these galaxies in large numbers across the full range of environments in the local Universe. These observations stress test simulations and challenge previous galaxy formation wisdom, with UDGs posing many open puzzles beyond just their unknown formation mechanism. To provide a few pertinent examples: there is observational evidence that not all UDGs follow the standard stellar mass\,--\,halo mass relationship; there is evidence for UDGs with extraordinarily high levels of alpha enhancement; and there is evidence that some UDGs are much more globular cluster rich than other dwarfs of similar stellar mass. In this Dawes review, we undertake the task of summarising the decade of science since the discovery of UDGs. We focus on the quiescent population of UDGs and review their general properties, their proposed formation scenarios, their internal properties and their globular cluster systems. We also provide a brief conjecture on some future directions for the next decade of UDG research. 
\end{abstract}


\maketitle

%%%%%%%%%%%%%%%%%%%%%%%%%%%%%%%%%%%%%%%%%%%%%%%%%%

%%%%%%%%%%%%%%%%% BODY OF PAPER %%%%%%%%%%%%%%%%%%

\section{Introduction}

{\it ``The Virgo Cluster survey has uncovered a new class of dwarfs that are of huge size  (10\,000\,pc in diameter in the extreme) and of very low surface brightness of about 25\,mag/arcsec$^2$ (B) at the centre''.} This is a quote from \cite{1984AJ.....89..919S} in their ground-breaking study of dwarf galaxies in the Virgo cluster. They identified around two dozen of this new class of kpc-sized, low surface brightness (LSB) galaxies. They went on to say, {\it `Most are in the central region of the cluster, suggesting that they are indeed members. No galaxies such as these low surface brightness, large dwarfs are known yet in the local neighbourhood'.} We now know that many such galaxies are indeed cluster members, and that local analogues also exist. 

For almost twenty years, these faint galaxies received little attention, {and were generally studied along with other galaxies of similar stellar mass (e.g., \citealp{2003AJ....125...66C}).} Based on their smooth appearance and large sizes in the Perseus cluster, \cite{2009AN....330..991P} argued that such galaxies must be dark matter dominated to prevent them from being disrupted by the cluster potential. {In some cases, the large size of these galaxies caused their explicit exclusion from catalogues of low surface brightness cluster members. For example, the study of \citet{Adami2006} searched the Coma cluster for ``fainter objects... with a radius larger than 0.6''} \footnote{{\citet{Adami2006} erroneously converts their 0.6'' radius to 3 kpc at Coma cluster distance. It should be 0.3~kpc. }}{... and with an R central surface brightness fainter than $\sim24$~mag arcsec$^{-2}$''. Here they applied a cut on the maximum radius of an LSB galaxy in their catalogue being 2.5'', which explicitly excludes any LSB galaxies larger than $\sim1.2$~kpc in radius, such as those uncovered by \citet{1984AJ.....89..919S}.}  

Later on, the interest in these large-sized LSB dwarf galaxies was rejuvenated by the discovery of {\it Forty-seven Milky Way-sized, Extremely Diffuse Galaxies in the Coma Cluster} by \cite{vanDokkum2015} using the new Dragonfly Telephoto Array. {It was unexpected to find that galaxies resembling those found by \citet{1984AJ.....89..919S} were so widespread in the Universe, with van Dokkum et al. noting only a handful of examples from previous studies \citep{Impey1988, Bothun1991, Dalcanton1997}.} As stated in the title of their article, these LSB galaxies have sizes (in terms of their semi-major effective radii, R$_{\rm e}$) comparable to the disk of the Milky Way. They also tended to be red, roundish and featureless. A working definition of R$_{\rm e}$ $\ge$ 1.5 kpc and central surface brightness $\mu_{g, 0}$ $\ge$ 24 mag/arcsec$^2$ in the $g$-band was established to broadly select the newly discovered galaxies. Both of these criteria are somewhat arbitrary and, as such, these LSB galaxies are not distinct from the general galaxy population in this parameter space \citep{Li2023c}. 
These galaxies were dubbed ``ultra-diffuse galaxies" (UDGs), and it was speculated that they were dark matter-dominated. {Some were proposed to be a ``failed galaxy'', existing in a region of stellar mass -- halo mass space where galaxies were previously not expected to reside}. Spectroscopic follow-up soon occurred, confirming their membership of the Coma cluster \citep{vanDokkum2015b} and their dark matter domination. Hubble Space Telescope (\textit{HST}) imaging then revealed the presence of rich globular cluster (GC) systems in many of them \citep{vanDokkum2017b}. 

\begin{figure*}
    \centering
    \includegraphics[width=0.95\textwidth]{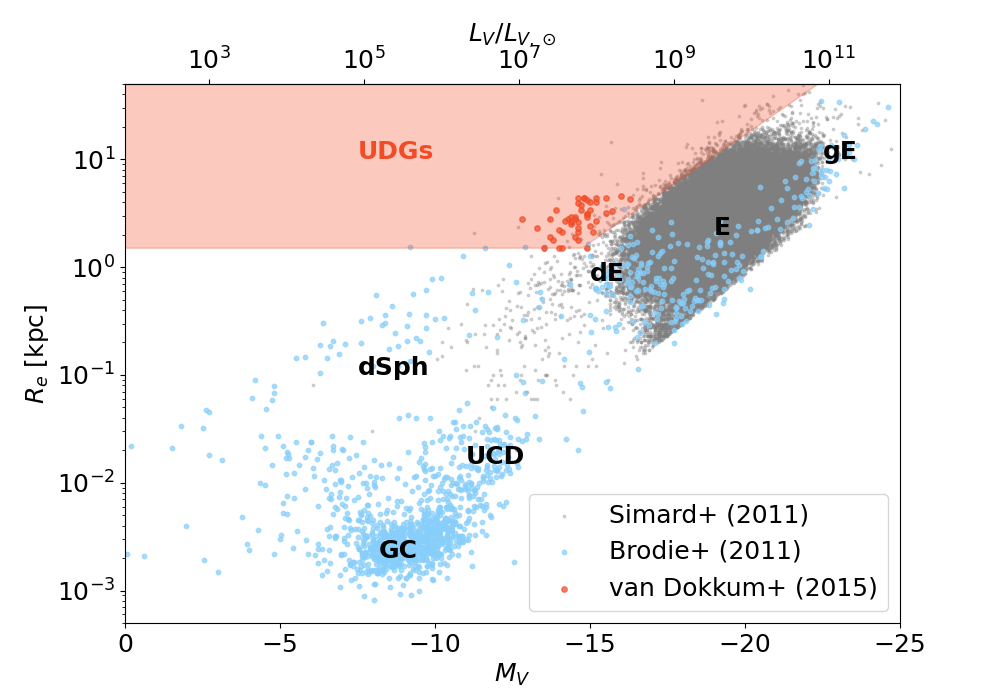}
    \caption{Half-light radius \textit{vs.} V-band magnitude for a range of (primarily elliptical) systems. Blue points are from \citet[and updates][]{Brodie2011}, grey points represent galaxies in SDSS with $0.01<z<0.05$ from \citet{Simard2011} and red points are the UDGs first detected in the Coma Cluster by \citet{vanDokkum2015}. The regions approximately corresponding to globular clusters (GC), dwarf spheroidals (dSph), ultra-compact dwarfs (UCD), dwarf ellipticals (dE), normal ellipticals (E) and giant ellipticals (gE) are as labelled. The region meeting the original UDG definition is shaded red and labelled appropriately. UDGs reside in a region of parameter space previously sparsely populated by surveys such as SDSS (highlighted in red).} 
    \label{fig:everything_plot}
\end{figure*}

It has now been 10 years since the resurgence in interest towards UDGs and LSB dwarfs. While hundreds of papers have been published on these galaxies in the last decade, there has yet to be a review. We undertake that task. In this review, our primary focus will be on the old and passive UDGs, i.e. galaxies whose stars formed some time ago and are not experiencing current star formation. {Unless such galaxies are observed during a quiescent period of cyclic star formation, or accrete gas at later times and recycle (e.g., \citealp{ascencio2025caught-051}),} they are expected to have a relatively stable effective radius and surface brightness, and thus remain a `settled' UDG. On the contrary, star-forming UDGs, despite being very interesting in their own right (e.g., \citealp{Leisman2017, ManceraPina2019b, Kong2022, Karunakaran2024}), represent a state that may be transitory given that they often contain HI gas and possible future star formation. 
%This type of UDG tends to be found in low density environments and often contain HI gas. While very interesting in their own right \citep{2019ApJ...883L..33M}, their UDG status may be transitionary. 
Our focus on old, passive UDGs implies an emphasis on UDGs located in high-density environments. As such, they tend to be red and gas-free. Some of them also have the interesting properties of hosting rich GC systems and perhaps occupying massive dark matter halos. Despite a decade of progress on UDGs, much remains to be explained. For brevity, we focus on the current state of the field and omit some of the historical context that has brought us to our current understanding.
%colour-environmental trend 
%\citep{Prole2019b}. 

{In Section \ref{sec:general_properties} we summarise UDGs' general photometric properties, including how they are defined, where they are located and their general morphological properties. In Section \ref{sec:formation_scenarios} we summarise the proposed formation scenarios of UDGs. We focus on providing short descriptions of each, along with a summary of testable predictions for each scenario. We then review the progress made studying the internal properties of UDG kinematics and dark matter content in Section \ref{sec:internal_properties} and their internal stellar populations in Section \ref{sec:stellar_pops}. UDG globular cluster systems are reviewed in Section \ref{sec:GCs}. In Section \ref{sec:twoclasses} we provide a brief comparison of UDGs to classical dwarf galaxies. We then summarise some possible future directions of UDG research in Section \ref{sec:future}. Conclusions are presented in Section \ref{sec:conclusions}. }

\section{General Photometric Properties} \label{sec:general_properties}
\subsection{UDGs in a New Region of Parameter Space}

The importance of \citet{vanDokkum2015}'s discovery was not necessarily that large, low surface brightness dwarfs existed, but instead that they existed in such large numbers. To illustrate their discovery, Figure \ref{fig:everything_plot} shows galaxies detected in SDSS with redshifts between $0.01<z<0.05$ from \citet{Simard2011} along with a large sample of pressure-supported systems from \citet{Brodie2011} that spans everything from compact globular clusters to giant ellipticals at the centre of galaxy clusters. UDGs from \citet{vanDokkum2015} are also included along with a shading for the original UDG definition assuming an approximate galaxy colour of $V=g-0.3$. While a handful of galaxies had been detected in SDSS in this region of parameter space, the addition of UDGs greatly increases the known population of galaxies with large half-light radii and $M_V\approx-15$. In addition to noting their large sizes in comparison to traditional dwarf ellipticals, it is useful for the remainder of this work to note the positioning of GCs and ultra-compact dwarfs (UCDs) in the lower centre of the plot. Many UDGs host significant populations of GCs and, in some cases, even UCDs. Understanding their connection to these compact systems is the subject of significant (ongoing) work and is described here in Section \ref{sec:GCs}. 

\begin{figure*}
    \centering
    \includegraphics[width=0.95\textwidth]{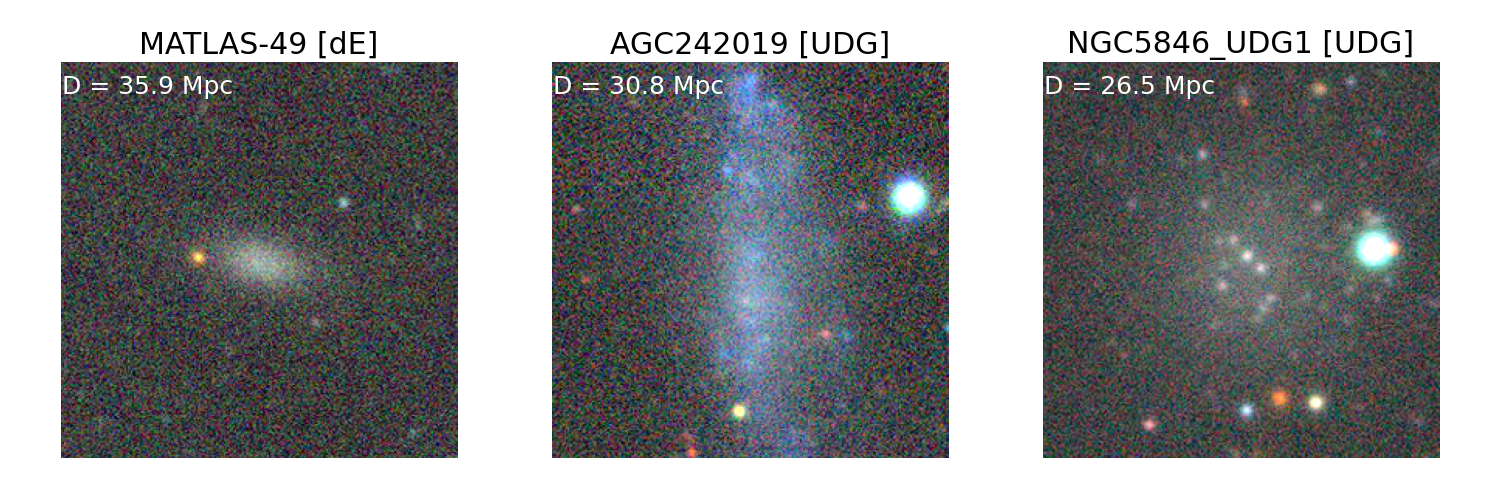}
    \caption{Dark Energy Camera Legacy Survey (DECaLS) images of three galaxies chosen for their similar distance as indicated in the top left of each cutout. The brightness of the images has been increased by a factor of 2 to aid the visibility of the UDGs. \textit{Left:} A prototypical dwarf elliptical, MATLAS-49. \textit{Centre:} An HI-bearing field UDG, AGC242019. \textit{Right:} A quiescent, GC-rich UDG, NGC5846\_UDG1 (MATLAS-2019). UDGs are noticeably larger than their dwarf elliptical (and dwarf irregular, although not shown here) counterparts. The UDG definition takes in a range of morphologies from blue, star-forming field UDGs to red, quenched group/cluster UDGs.}
    \label{fig:UDG_examples}
\end{figure*}

It is also of note that despite the initial UDG definition being coined for galaxies within the Coma cluster, which are strongly biased to being quiescent (see further Section \ref{sec:bias_quiescent}), galaxies meeting the UDG definition have now been found in all environments. In particular, UDGs have been found in significant quantities in the field (e.g., \citealp{Leisman2017}) where they may help to explain HI detections previously thought to reside in dark halos (i.e., those that did not form stars). UDGs found in the field are biased towards being bluer and more irregular in morphology. A significant fraction of field UDGs are found to be quiescent \citep{Prole2019b}, however the observed quiescent fraction is generally larger than, and in some tension with, simulations of dwarf galaxy formation \citep{Sales2022}. We provide examples of two UDGs in Figure \ref{fig:UDG_examples} along with a dwarf elliptical to provide scale for these galaxies. In particular, AGC242019 is a good example of an irregular, HI-rich, blue, star-forming field UDG, while NGC5846\_UDG1 (a.k.a. MATLAS-2019; N5846-156) is a good example of a quiescent, smooth morphology, GC-rich group UDG. Many of the compact sources visible on NGC5846\_UDG1 are indeed spectroscopically confirmed GCs 
\citep{Muller2020, 2025MNRAS.539..674H}. 
The three galaxies have been chosen for being located at approximately similar distances, to facilitate a fair visual comparison. 

\subsection{Alternative UDG Definitions} \label{sec:bias_quiescent}
It is worth noting that, following the initial definition of what comprises a UDG by  \citet{vanDokkum2015}, many subsequent authors have tweaked the definition for their own work. In general, alterations to the definition follow five main camps:
\begin{enumerate}
    \item Switching from a central surface brightness to one measured at, or the average within $R_{\rm e}$. Similarly, some simulations use the mean stellar surface density within 1 $R_{\rm e}$.
    \item Using a brighter surface brightness cut.
    \item Switching the filter band from the $g-$band to another e.g., $V-$~or $r-$band.
    \item Lowering the size cut from 1.5~kpc to e.g., 1~kpc. 
    \item Adding additional requirements, e.g., S\'ersic $n$~criteria or colour information.
\end{enumerate}

% Further, some authors call those UDGs that are close to the original UDG definition, nudging up against it, NUDGES \citep{Forbes2024}.

In the simulation work of \citet{vanNest2022}, the difficulties of comparing UDG samples defined via alternative definitions were explored. In particular, it was shown that more restrictive definitions as to what comprises a UDG lead both to fewer galaxies being selected as a UDG and those UDGs having a more preferred formation pathway compared to dwarf galaxies of a similar mass that do not meet the UDG definition. More restrictive definitions can also lead to galaxy orientation playing a role in selecting UDGs. In particular, as the definition is surface brightness-based, there is an orientation dependence as to whether or not a galaxy is defined as a UDG. That is, many UDGs that are viewed face-on will be brighter in surface brightness if viewed edge-on and would not be classified as a UDG \citep{Rong2020c}. This bias will also present as a bias in the ellipticity (axial ratio) of UDGs to be lower (higher) than those of the general dwarf population at the same stellar mass. There is likely an environmental dependence of this bias, where UDGs are diskier in the field than they are in clusters. For a full exploration of different UDG definitions and their limitations, see \citet{vanNest2022}.

Furthermore, it is well known that the UDG definition is generally biased against galaxies with young (bluer) stellar populations \citep{Li2023c, Li2023b}. For many galaxies with young stellar populations, their surface brightness will be transient, and as their stars evolve and die, they will fade into the UDG regime quickly \citep{Trujillo2017, Bellazzini2017, Roman2021}. Put another way, these blue galaxies exhibit UDG-like stellar surface densities, despite not meeting the surface brightness definition.

\citet{Li2023c, Li2023b} also pointed out that, due to the evolution in the size\,--\,stellar mass relationship with stellar mass, the initial UDG definition does not necessarily do a good job at selecting the largest galaxies at fixed stellar mass. Indeed, there is a clear change in the relationship of $R_{\rm e}$ with $M_{V}$ in Figure \ref{fig:everything_plot}. While we have shaded the UDG definition down to extremely low $M_V$, the size cut on the definition means galaxies of lower stellar mass than those initially discovered will also likely have to have extremely low surface brightnesses to make the size cut and are thus not readily observable beyond the Local Group. They propose it may be useful to study `ultra-puffy dwarfs', i.e., those that are outliers in the size-stellar mass relationship \citet{Li2023c, Li2023b}. Other studies have used this approach (e.g., \citealp{Lim2018}), and it is a common way to select `UDGs' in simulations with poor reproduction of dwarf galaxy sizes in general.

Finally, it is worth noting that the initial UDG definition left a small gap in parameter space between the galaxies usually studied in SDSS that tend to be brighter than the night sky ($\mu_{g, 0}<23$~mag arcsec$^{-2}$) and UDGs themselves ($\mu_{g,0}>24$~mag arcsec$^{-2}$). Rather than altering the UDG definition to include these galaxies, \citet{Forbes2024} have suggested these galaxies be dubbed NUDGEs (nearly UDGs) as they nudge up against the UDG definition. NUDGEs also include galaxies of slightly smaller size than UDGs but with a similar surface brightness (e.g., \citealp{FerreMateu2023}). For clarity in this work, we will endeavour to refer to UDGs meeting the original UDG definition as such, and will refer to NUDGEs when discussing galaxies that are close to, but do not strictly meet, the UDG definition.

\subsection{UDGs Numbers vs. Environment}
\begin{figure}
    \centering
    \includegraphics[width=0.45\textwidth]{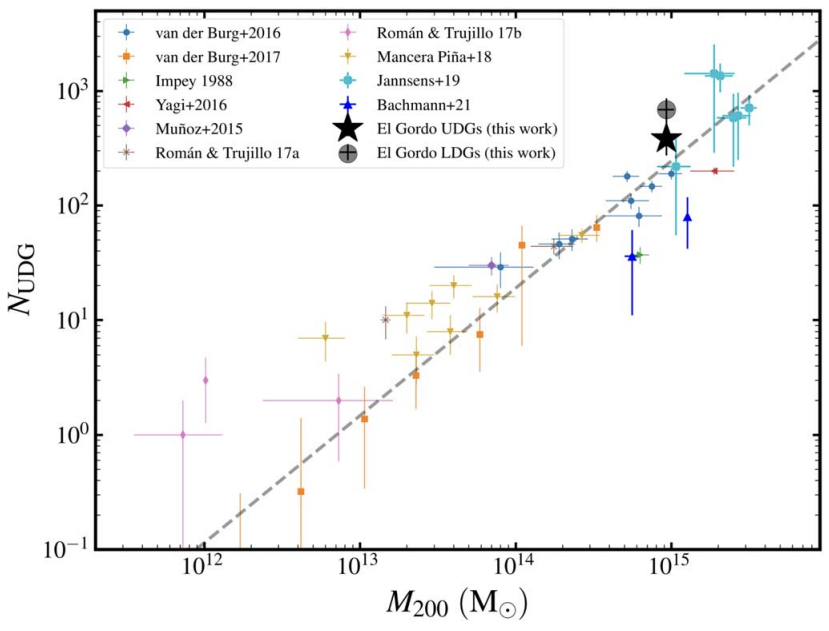}
    \caption{Number of UDGs \textit{vs.} total environmental halo mass, reproduced from \citet[][their figure 7]{Carleton2023}. {See also \citet{Karunakaran2023} and \citet{Goto2023} alternative plots of this relationship published at a similar time, but focusing on lower density environments.} Data are included from the studies of \citet{Impey1988, Munoz2015, vanderBurg2016, vanderBurg2017, Yagi2016, Trujillo2017, Roman2017b, ManceraPina2018, Janssens2019, Bachmann2021} per the legend. \citet{Carleton2023}'s work on El Gordo UDGs is indicated as ``this work" in the legend. The dashed line represents the relationship derived in \citet{vanderBurg2017} of $N_{\rm UDG} \propto M_{200}^{0.93\pm0.16}$. The near linearity of this relationship indicates that the environment may not play a strong role in the net creation/destruction of UDGs.}
    \label{fig:nudg_m200}
\end{figure}

The most basic questions to answer for UDGs are simply: \textit{How many of them are there? And where are they all?} These questions are important for two main reasons. i) If UDGs do not represent a significant fraction of the galaxy population, much of their formation could be ascribed to them being many sigma outliers in established and well-explored processes of galaxy formation. While they would still be worthy of study, their existence would perhaps not represent a large challenge for galaxy formation. ii) Knowing where UDGs are found is key to understanding the role that their environment plays in their formation. If UDGs are found only in massive galaxy clusters, it suggests that environmental processes play a strong role in their formation and evolution. Conversely, if UDGs are only found in the field, they must form via internal processes, and they may be efficiently destroyed once they enter a dense environment. 

In Figure \ref{fig:nudg_m200} we reproduce figure 7 of \citet{Carleton2023}, which shows the number of UDGs detected as a function of total environmental halo mass (a similar, recent plot can be found in 
\citealp{Karunakaran2023}). As shown in Figure \ref{fig:nudg_m200}, UDGs have been detected in all environments from the field (e.g., \citealp{Leisman2017}) to the most massive clusters in the nearby Universe (e.g., \citealp{Janssens2019}). There still exists some debate around the exact slope of the relationship between the number of UDGs ($N_{\rm UDG}$) and the halo mass ($M_{200}$); however, most studies agree that its slope is approximately 1. Slopes less than unity would imply that dense environments do not promote UDG formation and likely even play a role in destroying UDGs as they fall in from the field. Slopes greater than unity imply the opposite, that the external processes operating in groups/clusters play an important role. As such, current results indicate that the environment may not play a strong role in UDG formation/destruction or, at the very least, any increase in UDG production caused by dense environments is being roughly balanced by the UDG destruction in them. 

Perhaps the greatest problem with our current understanding of UDG formation's relationship with the environment is simply the different UDG definitions being used by the many authors plotted in Figure~\ref{fig:nudg_m200}. Having different definitions for what a UDG is when calculating their number will clearly affect the relative positions of different studies to one another and thus the slope of the relationship between $N_{\rm UDG}$~and $M_{200}$. A further consideration is the correction to UDG counts out to the virial radius. Imaging usually covers only a fraction of the cluster, creating the need to correct the UDGs detected in the imaged area to estimate the total number in the cluster. Different studies have employed varying approaches to make this correction. Without a systematic study of UDGs in different environments using the same definition and the same corrections, it is not possible to quantify the effects these issues have on our current ability to determine the slope of the $N_{\rm UDG}-M_{200}$ relation in different environments.  

% For example: \citet{vanderBurg2016} used a UDG definition of $24\leq \langle \mu_r \rangle_e \leq 26.5$~mag arcsec$^{-2}$, $1.5\leq R_{e} \leq 7$~kpc and S\'ersic index $n\leq 4$; While \citet{Janssens2019} used a UDG definition of $24.1\leq \langle \mu_{F814W} \rangle_e \leq 26.9$~mag arcsec$^{-2}$, $1.5\leq R_{e} \leq 10$~kpc,  S\'ersic index $n\leq 4$ and axis ratio $b/a>0.3$. 

The answer to the question regarding what fraction of the general galaxy population are UDGs appears to be approximately 5\% of all galaxies with stellar mass $\gtrapprox 10^7~M_\odot$ across all environments \citep{Jones2018}. For example, \citet{Li2023c} find that $7\pm2$ \%~of the satellites around systems that are analogous to the Milky Way are UDGs. A similar expectation from simulations is seen in \citet{Newton2023}. Due to the linear slope found in Figure \ref{fig:nudg_m200}, UDGs being approximately 5\% of all galaxies remains true in the most massive clusters. As a confirmation, the work of \citet{Janssens2019} found 1351$^{+387}_{-379}$ UDGs to reside in the Hubble Frontier Field galaxy cluster Abell 2744. With the \textit{JWST} UNCOVER survey recently discovering $\sim50000$ sources of near-infrared light in the vicinity of that cluster \citep{Weaver2024}, these UDGs again represent a significant galaxy population within the cluster ($\sim3$\%). Thus, any works based on galaxy mass/luminosity functions must take into account UDGs. {A consequence of the abundant nature of UDGs is that they outnumber L$^\star$~galaxies of similar half-light radius \citep{Danieli2019b}.}

\subsection{The Morphology of UDGs}

UDGs exhibit a wide range of morphologies, but {the red, quiescent UDGs reviewed herein} tend to be largely featureless beyond hosting GCs and nuclear star clusters. There still exists some debate if they are oblate or prolate in nature (see \citealp{Burkert2017} and \citealp{Rong2020}). Their surface brightness profiles are typically exponential disk-like, with mean S\'ersic indices of $n\sim0.8$ \citep{Mowla2017}. In some cases, disk-like profiles are used as a selection criterion. Evidence for traditional features such as bars, arms or bulges are rare. Tidal features are seen in some cases, but the majority of UDGs appear to be relatively undisturbed. For example, in the recent visual analysis of 7000 UDG candidates in the SMUDGes survey, some 93\% were classified as `undisturbed UDG candidates' \citep{2025OJAp....8E..90Z}. 
Earlier work by \cite{Mowla2017}, stacking deep images of 287 UDGs, found no evidence of tidal features out to $\sim$4 R$_{\rm e}$. This suggests that if they did form by merger/interaction, then it was sufficiently long ago so that signatures are no longer seen. 

The MATLAS survey found around 1/3 of UDGs to host a nuclear star cluster \citep{Marleau2021, Poulain2025}. For the SMUDGES survey, {of slightly shallower depth}, the number was more like 1/10 \citep{Lambert2024}. {We emphasise that} these fractions are comparable to those for the classical dwarfs in the same survey, but slightly lower than the fractions observed in high-density environments such as the Coma cluster. For the stellar masses of UDGs, it is expected that nuclear star clusters are the result of GC infall and merging via dynamical friction \citep{2021A&A...650A.137F}. \cite{Janssens2019} presented evidence from the Frontier Field clusters that UCDs may be the nuclear remnants of stripped UDGs that contained a nuclear star cluster. This concept was further supported by the diffuse envelopes seen around some Virgo cluster UCDs by \cite{Wang2023}.

\section{UDG Formation Scenarios and their Simulations} \label{sec:formation_scenarios}
Reproducing the observed population of dwarf galaxies in simulations within the $\Lambda$CDM paradigm is an extremely difficult problem due to the necessity of including sufficient physics (e.g., feedback effects/ISM physics), sufficient volume (e.g., for environmental effects) and sufficient resolution (e.g., to resolve the dwarfs) to accurately model their properties (\citealp{Sales2022}). 
Before the discovery of UDGs there was an established problem with simulations whereby they were producing large numbers of dwarf galaxies that were generally larger than those being observed (see e.g., \citealp{Sales2022}). The discovery of UDGs showed that some of these galaxies may exist. Their discovery, therefore, represented a post-diction (sic.) of simulations. Key to the simulation of UDGs, is the need to simultaneously create normal dwarfs and to demonstrate what sets UDGs apart in their formation pathway(s). Generally, the proposed formation mechanisms in simulations are split into those that occur in isolation within a halo (i.e., internal formation) and those that occur due to the influence of external halos (i.e., external formation). It is also established that galaxies may not need to be transformed into a UDG by external or internal processes, but may instead form intrinsically as a UDG, i.e. born as a UDG.

We next discuss some of the major attempts to elucidate UDG formation pathways, and include some testable predictions expected in each scenario. 

\subsection{Internal Formation}
\subsubsection{High Halo Spin}
\textit{Description:} In this formation scenario, the UDG forms as a normal dwarf galaxy, although its dark matter halo has above average spin. This additional angular momentum within the dark matter halo causes it to be more diffuse, which, in turn, results in a more diffuse galaxy (i.e., the UDG). For example, \citet{Jiang2019} found that a galaxy's half-light radius is proportional to its halo spin along with its virial radius. It is of note that high halo spin may also be correlated with a bias to later halo assembly. This scenario is primarily simulated/modelled in \citet{Amorisco2016, Rong2017} and \citet{Liao2019}.  

\textit{Testable Predictions:} As noted in \citet{Amorisco2016}, the halo spin distribution is not considered strongly environmentally dependent, so this formation scenario is expected to function in all environments. For instance, \citet{Liao2019} found that the number of UDGs follows an approximately linear power law with environmental mass, while \citet{Amorisco2016} provided a prediction for the number of UDGs expected as a function of their luminosity. It is unclear if the above-average angular momentum of the dark matter halo also translates into above-average angular momentum within the galaxies. {In general, galaxy halo spin has not been found to correlate strongly with galaxy angular momentum within cosmological simulations \citep{Jiang2019b}.} However, \citet{Rong2017} predicted that in the field, where environmental effects are minimal, UDGs should be on average redder than classical dwarfs. This is due to their more extended nature, causing their gas densities to drop below the threshold for star formation earlier, resulting in on average older stellar populations. Finally, \citet{Rong2017} predicted that UDGs forming via this pathway should have, on average, later cluster infall times than their dwarf counterparts (median population infall times of 9\,Gyr vs 5\,Gyr in time since the Big Bang).

\subsubsection{Supernova Feedback}
\textit{Description:} In this formation scenario, UDGs form as a classical dwarf galaxy, but intense, repeated, bursty episodes of star formation (and thus star formation feedback) help to redistribute mass away from the centre of the halo. This will both core the dark matter halo and cause an increase in the size of the dwarf galaxy it hosts, lowering its stellar surface density and causing it to become a UDG. It is unclear what causes the stronger stellar feedback needed to puff up a UDG over the regular feedback experienced by a normal classical dwarf of similar stellar mass. This scenario is primarily simulated in \citet{DiCintio2017} and \citet{Chan2018}, with both simulating UDGs forming in isolation. 

\textit{Testable Predictions:} Due to the need for repeated episodes of star formation, and the fact that ejected gas will have re-accretion times significantly less than a Hubble time, the star formation histories of UDGs forming in this way are expected to be rather bursty. This may result in a (temporarily) quenched UDG surrounded by ejected gas. These UDGs are also expected to have cored dark matter halos. The maximum halo mass of UDGs forming via this mechanism is expected to be roughly $10^{11}~M_\odot$~\citep{DiCintio2017}. The higher the half-light radius, the higher the fraction of young stars are expected to be found within isolated UDGs forming via this mechanism \citep{DiCintio2017}. A possible mix of radial age and metallicity gradients from flat to negative is expected for these galaxies \citep{Chan2018, CardonaBarrero2022}. 
 
\subsubsection{High Gas Spin}
\textit{Description:} In this formation scenario, the proto-UDG exhibits higher than average energy within its star-forming gas disk throughout its life. As the galaxy forms, this gas disk is therefore more extended and, due to the extra angular momentum, takes longer timescales to collapse, causing less efficient star formation. The natural effect of this is a larger, more diffuse dwarf galaxy (i.e., a UDG). It is noteworthy that it is unclear if this is true for the entire gas disk or just for the star-forming portion of the gas. This scenario is primarily simulated in \citet{Zheng2025}.

\textit{Testable Predictions:} A clear prediction of this scenario is that the gas within these dark matter halos should exhibit both a larger angular momentum and a greater extent than normal dwarf galaxies \citep{Zheng2025}.

\subsection{External Formation}
\subsubsection{Tidal Heating/Puffing}
\textit{Description:} In this formation scenario, energy is injected into the stellar body of a regular dwarf galaxy through the gravitational potential via (likely repeated) close passages with more massive halos, increasing its size and lowering its surface brightness. {It is possible this formation scenario represents an extension of the ``galaxy harassment'' model previously proposed to turn spiral galaxies into dwarf elliptical/spheroidal population within clusters \citep{Moore1996}.} It is likely related to the formation scenario described below for UDGs forming via tidal stripping. It is unclear if this formation scenario is more efficient in cored dark matter halos given the differing results from \citet{Carleton2019} and \citet{Sales2020}. Further, it is clear that this scenario likely applies to many UDGs that have fallen into cluster/group environments from the field (see e.g., \citealp{Jiang2019}). This scenario is primarily simulated in \citet{Carleton2019, Jiang2019, Liao2019} and \citet{Sales2020}.

\textit{Testable Predictions:} This scenario requires a more massive halo in order to work, thus UDGs are unlikely to form via this scenario in the field. Due to the repeated passages required to make the largest UDGs ($R_{\rm e}>3$\,kpc), these are expected to host old stellar populations (89\% $>$4\,Gyr; \citealp{Carleton2019}). UDGs should show signs of dark matter loss in their outskirts, concentrating the dark matter distribution in the centre. UDG abundances should be correlated with cluster age and drop off with redshift \citep{Carleton2019}. UDGs formed via tides in this manner should experience significantly earlier infall times ($\sim$9.5\,Gyr ago) to the cluster than those that have formed as UDGs prior to cluster infall ($\sim$5.5\,Gyr ago). 

\subsubsection{Tidally Stripped Galaxies}
\textit{Description:} In this formation scenario, the UDG represents the remnant of a more massive galaxy that has experienced significant tidal stripping of stars to reduce its stellar mass into the dwarf regime. More moderate tidal stripping is possible to aid in the formation of a UDG, however, other mechanisms are likely needed. This scenario is primarily simulated in \citet{Sales2020} using the disruption models of \citet{Errani2015}. %Their resulting UDGs are named T-UDGs. 
This is also simulated as part of the work of \citet{Carleton2019}.

\textit{Testable Predictions:} They should be metal-rich, reflective of their origin as a more massive and now disrupted galaxy \citep{Sales2020, Benavides2024}. They should also be relatively dark matter-free, as any tidal interaction able to strip significant stellar mass from a galaxy will also have stripped much of their dark matter halos \citep{Sales2020}. Finally, tidal features are expected for these galaxies, indicative of their tidal stripping. \citet{Carleton2019} found that S-shaped tidal features are expected to fade within $\sim0.7$\,Gyr of creation and that $\sim13\%$ of their UDGs have experienced a pericentric passage within this time. Thus, approximately $13\%$ of UDGs forming via tidal stripping should display tidal features in clusters \citep{Carleton2019}.

\subsubsection{Galaxy Mergers}
\textit{Description:} In this formation scenario, the UDG forms via a major merger of two smaller dwarf galaxies. In the scenario described by \citet{Wright2021} the UDG forms via a high redshift merger at $z>1$. These mergers temporarily redistribute star formation to the outskirts of the galaxy, causing its large size. As the galaxy ages, its surface brightness fades due to passive stellar evolution. 

\textit{Testable Predictions:} As the spin up caused by the merger occurs at high-$z$, many of the distinguishing features of this scenario are temporary and are not visible at $z=0$ \citep{Wright2021}. However it is an expectation that $\sim20\%$ of dwarf galaxies in the stellar mass range of $10^7 - 10^9 M_\odot$ form via this pathway. They do so in exclusively dwarf like halos of mass $<10^{11} M_\odot$~\citep{Wright2021}. In scenarios other than \citet{Wright2021}'s where UDGs form via galaxy mergers, prolate rotation may be expected as is commonly seen in more massive galaxies formed via mergers. Finally, if the merger were wet we would expect a strong burst in their star formation history at the time of the merger.

\subsubsection{Ram Pressure Stripped Galaxies}
\textit{Description}: In this formation scenario, a dwarf galaxy falling into a cluster has its gas ram-pressure stripped, which quenches star formation. Passive evolution will then lower the surface brightness into the UDG regime. Further, the removal of gas from the centre of the halo will flatten the gravitational potential and may help the galaxy to expand, becoming a UDG. This scenario is primarily discussed in: \citet{Yozin2015, Chilingarian2019, Grishin2021} and \citet{Junais2022}.   

\textit{Testable Predictions}: Due to the contraction of star-forming region during a starburst, as the gas is stripped, it is expected that an age gradient will form within the proto-UDG as it is stripped of gas \citep{Grishin2021}. Likewise, a strong truncation of star formation in the star formation history of the galaxy is expected. Finally, it should be noted that it may take up to 10\,Gyr for the galaxy to fade into the UDG regime and thus the accretion time has to be relatively ancient for the galaxy to be observed to be a UDG today. It is unclear if there existed environments sufficiently massive 10 Gyr ago to cause similar stripping to what is seen today in clusters. Up to $\sim 45\%$ of UDGs in clusters have been claimed to form via this pathway \citep{Grishin2021}. 

\subsection{`Born' as a UDG}
\subsubsection{Tidally Stripped Stars}
\textit{Description:} In this formation scenario, the UDG forms from stellar material stripped from a more massive galaxy as it undergoes a tidal interaction. In this regard, it is just the large size--low surface brightness end of normal tidal dwarf galaxies \citep{Duc2012}. A key difference with the above \textit{Tidally Stripped Galaxy} scenario is that here the galaxy does not initially form within a dark matter halo; instead, it forms from stellar material stripped out of its host dark matter halo via a tidal interaction. 

\textit{Testable Predictions:} Predictions that are generally applicable to tidally stripped galaxies are likely also applicable to these galaxies. Namely, as they have formed from stars stripped from a more massive galaxy, their stars should have similar ages and be more metal-rich than is generally expected for a dwarf. They should also be dark matter-free, reflective of their formation without a dark matter halo. These UDGs are likely to be found embedded within larger tidal features and may host GCs if these were stripped along with the stars. In general, significant GC populations are not expected. In general, UDGs forming via this pathway are not expected to survive for long periods of time due to their lack of a dark matter halo and formation near a more massive galaxy. According to the simulations of \citet{Moreno2022}, roughly one third of massive galaxies may have a dark matter-deficient satellite (not necessarily a UDG).

\subsubsection{Tidally Stripped Gas}
\textit{Description: } In this formation scenario, the UDG forms in gas stripped from a more massive halo. The gas may be stripped via tidal interaction or ram pressure processes. UDGs forming in the ram pressure stripped gas of ``jellyfish galaxies'' falling into clusters fit into this scenario. It is slightly different from the previous scenario for \textit{``Tidally Stripped Stars''}, whereby the star formation occurs after the gas is stripped from the galaxy, rather than the stars themselves being stripped from the galaxy to form the UDG. This scenario is primarily described in \citet{Poggianti2019} and \citet{Ivleva2024}.

\textit{Testable Predictions:} As these UDGs form from tidally stripped gas, they are not expected to have their own dark matter halo and should have kinematics reflective of this. Likewise, as they form from gas stripped from a more massive galaxy, they should exhibit higher stellar metallicities than other dwarf galaxies of similar stellar mass. In general, it may be difficult to distinguish this scenario from the \textit{``Tidally Stripped Stars''} one without an indication of both the stellar age of the UDG and the age of the interaction during which its material may have been stripped.

\subsubsection{Bullet Dwarf}
\textit{Description:} This scenario represents somewhat of a combination of the above two scenarios. Namely, a string of galaxies, many of which may be UDGs, forms due to the high-velocity collision of two dwarf galaxies. It has been dubbed the ``bullet dwarf'' scenario \citep{Silk2019, vanDokkum2022} due to its resemblance to a small-scale version of the Bullet Cluster. Here, the collision separates the collisional (i.e., the gas) from the non-collisional (i.e., stars, star clusters and dark matter) elements of a dwarf galaxy halo, forming a line of new galaxy formation along the trail as the stripped gas collapses. Some of these galaxies may be UDGs. This scenario is primarily simulated/described in \citet{Silk2019, Shin2021, Lee2021, Lee2024} and \citet{vanDokkum2022}. 

\textit{Testable Predictions:} This scenario will present as a linear trail of galaxies on the sky, assuming that they have not had their orbits perturbed since formation. At either end of the trail should be the initial galaxies from the collision, being dark matter-rich and stellar mass-poor (reflecting their loss of gas), while galaxies between them should be dark matter-poor with similar stellar populations \citep{vanDokkum2022}. They may also exhibit above-average star cluster formation (both in terms of their absolute number and their individual masses), reflective of the higher gas pressures generated by such a collision \citep{Lee2021}. These galaxies may also exhibit stellar bodies stretching along the line of the collision \citep{Tang2025a}, along with prolate rotation \citep{Buzzo2025b}, as a memory of their formation with momentum from the collision. 

\begin{figure*}
    \centering
    \includegraphics[width=0.95\textwidth]{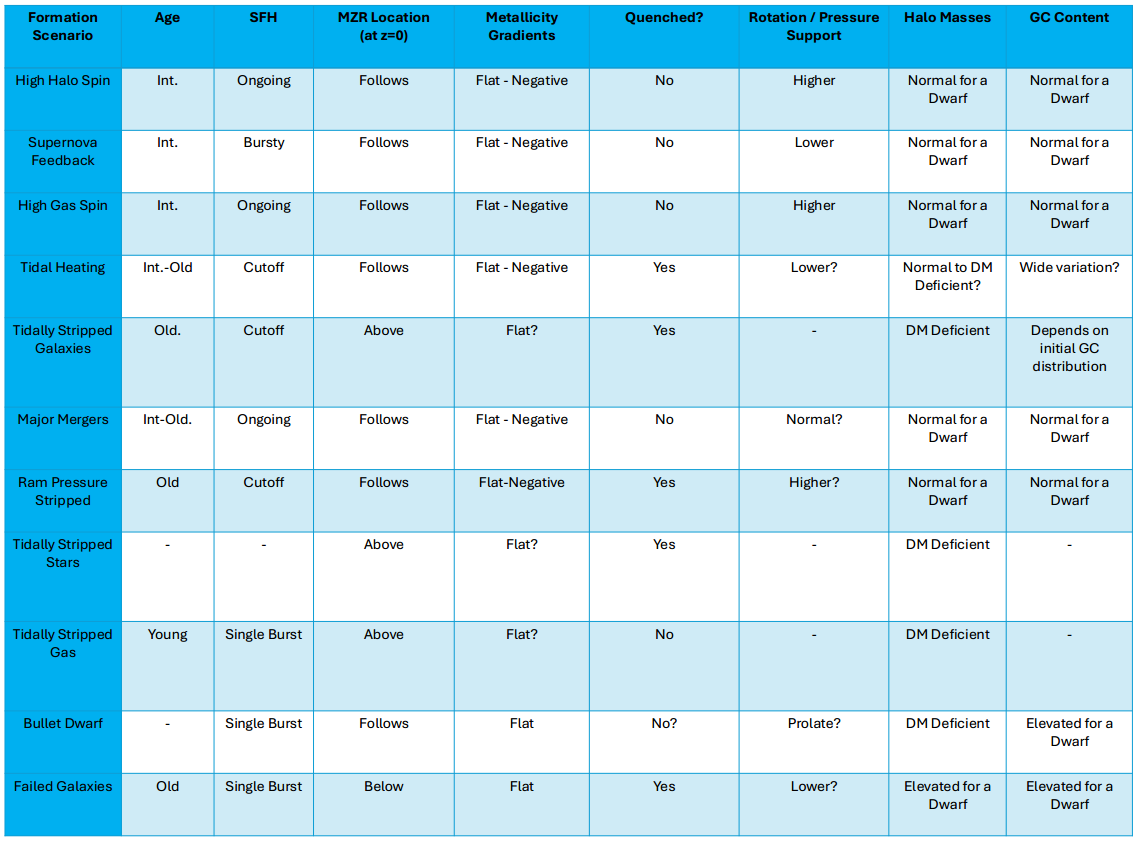}
    \caption{Summary table of predicted UDG properties for each formation scenario as described in section \ref{sec:formation_scenarios}. We either write `-' if the prediction is totally unclear or add a `?' next to a likely property that is yet not well-established. `Int.' refers to intermediate ages in first column. We provide this table as a basic guide for future UDG studies.}
    \label{fig:glance_table}
\end{figure*}

\subsubsection{Failed Galaxies}
\textit{Description:} In this formation scenario, the proto-UDG forms quickly at high redshift from a single burst of high-intensity star formation in a massive dark matter halo and is rapidly quenched. This high-intensity star formation drives the efficient production of GCs, which comprise most of the proto-UDG at high redshift. Beyond environmental quenching, there is no clear mechanism to quench, and keep quenched, the proto-UDG although feedback from the formation of the GC system has been proposed \citep{Gannon2025}. The proto-UDG then evolves passively for the remaining age of the Universe, where GCs slowly disrupt/evaporate to form/contribute to the stellar body of the galaxy and where stellar evolution causes the proto-UDG to slowly decrease in surface brightness into the UDG regime. It is dubbed a ``failed galaxy'' as its host dark matter halo fails to form the stellar mass it is expected to given a standard stellar mass\,--\,halo mass relation. This scenario has also been dubbed a ``pure stellar halo'' as this formation event is the epoch at which most galaxies form their stellar halo, and the UDG experiences no subsequent evolution \citep{Peng2016}. It is important to note that while this was initially dubbed a ``failed Milky Way galaxy'' scenario by \citep{vanDokkum2015}, there is now strong evidence that UDGs do not inhabit halos that are that massive \citep{ vanDokkum2019b, Gannon2020} but are more like M33. {Indeed, weak lensing analysis on stacked UDGs indicates that only a small fraction of UDGs could inhabit the relatively massive halo required for this scenario \citep{Sifon2018}.} This scenario is primarily described in \citet{vanDokkum2015, Peng2016, Beasley2016, Danieli2022, Forbes2024, Forbes2025} and \citet{Gannon2025}. 

\textit{Testable Predictions:} Due to the extreme nature of this formation scenario, it makes a number of testable predictions. Namely, the UDG should have: 1) an old age, elevated abundance of $\alpha$ elements, flat age and metallicity gradients, and (early) single burst star formation history; % reflective of its single burst of star formation, 
2) GC-like ages and metallicities reflective of its stellar body being significantly comprised of disrupted GCs; 3) a rich GC system reflective of its early, high-intensity star formation, and 4) a massive dark matter halo as required for it to be ``failed''. Additional predictions may be possible depending on the mechanism of their quenching. e.g., if the ``failed galaxies'' were environmentally quenched at high redshift, they are expected to be amongst some of the earliest dwarf galaxies to accrete onto the clusters they now observed to reside in. It is worth noting that \textit{JWST} has already revealed systems with high fractions of their stars within their GC system at high redshift \citep{Mowla2022, Mowla2024, Vanzella2022a, Vanzella2022b, Vanzella2023, Adamo2024, Messa2024, Bradac2025, Naidu2025}, proving possible examples of what a proto-failed galaxy-UDG may look like.

\subsection{UDG Formation Summary}
It is becoming increasingly obvious that UDGs cannot be fully described by 
any one of these formation scenarios alone, and that they provide a challenging `stress test' to models of galaxy formation. Several processes may be contributing to the overall UDG population. For example, there are some clear cases of UDGs forming that are associated with \textit{Tidally Stripped Stars} (e.g., CenA DWIII; \citealp{Crnojevic2017}) and \textit{Tidally Stripped Gas} (e.g., Hydra I-UDG32; \citealp{Hartke2025}). Furthermore, most proposed models cannot easily account for the UDGs with an elevated GC richness, beyond that of a classical dwarf. As described below in Section \ref{sec:stellar_pops}, SED and spectroscopic studies also point to the need for multiple formation pathways to explain the variety in their measured stellar populations and other characteristics (e.g., \citealp{Kadowaki2017, RuizLara2018, FerreMateu2018, FerreMateu2023, Buzzo2025}). In many cases, combinations of formation scenarios, e.g., their formation in the field and subsequent quenching/tidal heating by accretion onto a dense environment \citep{Roman2017b, Sales2020}, also seem likely.  

Even beyond the above-discussed scenarios, it is clear that other effects likely influence the formation and evolution of UDGs. For example, the ram pressure stripping of their gas in a single passage through a dense environment may cause quiescent UDGs to appear in the field as ``backsplash'' galaxies \citep{Benavides2021}. With all these different caveats in mind, we have created a reference table to summarise the basic predictions of each scenario as Figure \ref{fig:glance_table}. We stress that it should not be used in isolation from the detailed studies of each formation scenario, as summarised above. It is merely provided to guide future studies of UDGs. In particular, the column of `Halo Masses' can classify scenarios into those thought of as ``puffy dwarfs'' and ``failed galaxies''. That is, UDGs forming in a halo of mass normal for a dwarf may simply be the large size end of the normal dwarf size distribution, i.e., a ``puffy dwarf". Alternatively, UDGs forming in more massive halos than are typical for a dwarf fit the ``failed galaxy'' scenario. Finally, there are a number of scenarios of UDGs forming with dark matter deficiency. These tend to form from interactions with other systems.  

%\section{UDG Internal Properties} \label{sec:internal_properties}
\section{UDG Internal Properties I: Clues from galaxy kinematics} \label{sec:internal_properties}
Observationally, obtaining deep data to differentiate UDG formation pathways has been one of the major challenges of this decade. While deep imaging has been able to provide large samples of UDGs (e.g., the SMUDGES Survey; \citealp{Zaritsky2019}), many of the predictions of the formation scenarios (UDG halo masses/metallicities) require deep spectroscopy to test. From deep spectra, we can measure the galaxy's stellar kinematics (rotational velocities, velocity dispersions), which are crucial to obtain dynamical masses. The latter, under various assumptions, may provide an estimate of their total halo mass. Good quality spectra can also be used to probe the stellar populations of UDGs. This can provide better constrained ages and metallicities compared to SED fitting or colours, while also giving star formation histories (SFHs), quenching timescales, and other chemical signatures such as elemental abundance patterns. Currently there are $<$50 UDGs with their internal properties measured with spectroscopy \citep{Gannon2024}. In this Section (\ref{sec:internal_properties}; Clues from galaxy kinematics) and the following Section (\ref{sec:stellar_pops}; Clues from stellar populations), we elucidate what has been learned from these efforts. 

\subsection{What drives UDG dark matter content - Velocity Dispersions?}
Given the $L^\star$-like, large half-light radii but low, dwarf galaxy-like stellar masses, an initial question of their study was simply which galaxy class, $L^\star$~or dwarf, their dark matter halo resembled. Before getting too involved in the nuanced answer to this question, it is worth stating that there is very little evidence for any UDG residing in a $L^\star$-like dark matter halo. The weak lensing study of \citet{Sifon2018} placed a 95\% confidence interval upper limit on the halo masses of UDGs $\log( M_{200}/\mathrm{M}_\odot)\leq11.8$, which is less than typical of any $L^\star$-like galaxy (i.e., $\log( M_{200}/\mathrm{M}_\odot)\approx12$). Furthermore, most UDG dynamical masses are too small to be reasonably fitted by a massive, $L^\star$~dark matter halo \citep{Gannon2020}. For this reason, the field has reframed the original question as to whether or not UDGs reside in $L^\star$~or dwarf-like dark matter halos into whether or not they reside in dwarf-like dark matter halos, or those that are more massive than is typical for a dwarf. This is particularly relevant when discussing the ``failed galaxy'' scenario described above.

\begin{figure}
    \centering
    \includegraphics[width=0.45\textwidth]{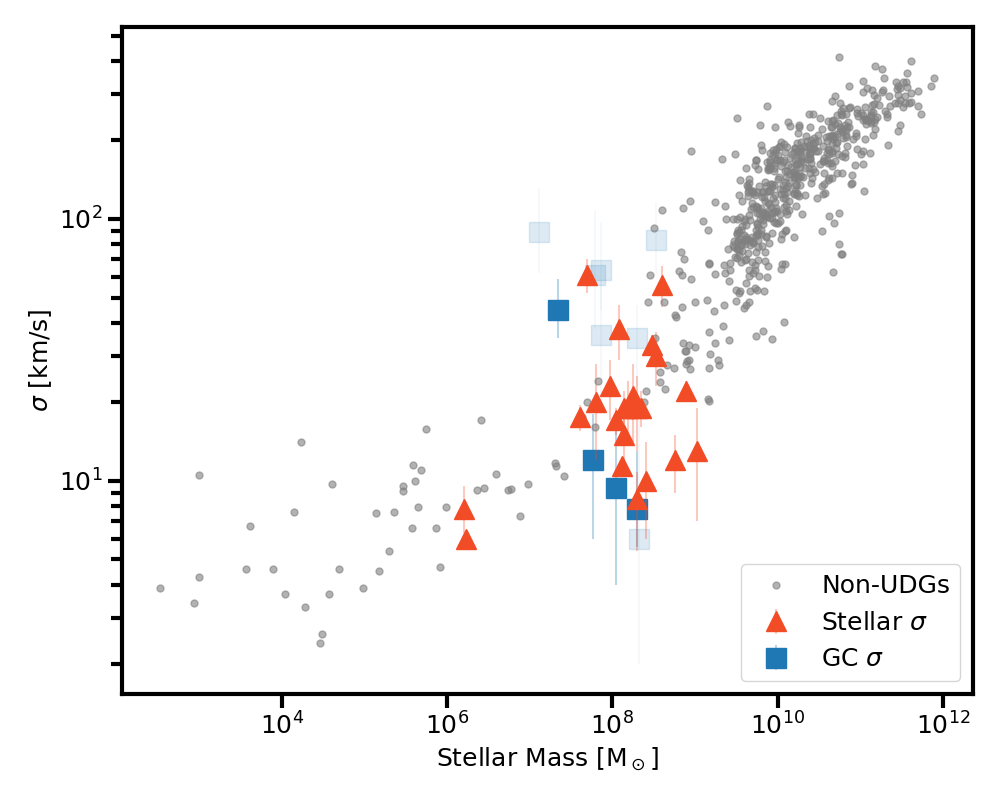}
    \caption{Galaxy velocity dispersion vs galaxy stellar mass. The non-UDGs that establish the normal relation are from the works of \citet[][grey points]{Chilingarian2009, mcconnachie2012, Cappellari2013, Harris2013}. UDGs are plotted from the catalogue of \citet{Gannon2024b} with dynamical masses coming from their stellar velocity dispersions (red triangles) and from their GC system velocity dispersions (blue squares). In general, UDGs with secure velocity dispersion measurements do not appear to be dynamically hotter or colder than dwarf galaxies of a similar stellar mass.}
    \label{fig:sigma_mstar}
\end{figure}

In order to answer this question we need to measure masses for UDGs. A common way to do this is to measure a velocity dispersion from the stars within the galaxy or from the dispersion of GC recessional velocities about the galaxy. When estimating dynamical masses for UDGs, the vast majority of studies have used the Jeans mass estimator of \citet{Wolf2010}, which is reliant on the square of galaxy's velocity dispersion along with its half-light radius. 

It is thus useful to investigate UDG velocity dispersions to see if they are typical for a galaxy of their stellar mass. In Figure \ref{fig:sigma_mstar} we show UDG velocity dispersions that are measured either internally from their stellar body or from the dispersion of their GC system vs. their stellar masses. When GC velocity dispersions are inferred from $<10$~GCs, we employ a greater transparency on these data points. Non-UDGs are also shown to establish the expected relationship for galaxies. The vast majority of UDG stellar velocity dispersions lie within the scatter of those for non-UDGs. 
 
There is an interesting population of four UDGs (Hydra-I UDG4, NGC1052-DF2, PUDG\_R15 and PUDG\_R16) in Figure \ref{fig:sigma_mstar} that have stellar velocity dispersions significantly below what is established for non-UDGs. The integrated velocity dispersion includes any rotation \citep{Corteau2014}. However, this rotational term will not include any inclination correction. As per the discussion in Section \ref{sec:general_properties}, the UDG definition is biased to selecting face-on galaxies, so this rotation correction may be significant in some cases. After the inclusion of this correction, these UDGs mentioned above may present more regular (higher) velocity dispersions. It is worth noting that multiple UDGs, along with a number of NUDGes, have recently been reported to exhibit varying degrees of rotation, suggesting this solution may be viable \citep{Chilingarian2019, Buttitta2025, Levitskiy2025}. Further, it is worth noting that a pair of NUDGes that did not meet the criteria to be a UDG used by \citet{Gannon2024b} but that have been referred to UDGs by some authors (FCC~224 and NGC1052-DF4; \citealp{vanDokkum2019, Shen2023, Tang2025b, Buzzo2025b}) also would reside in a similar region to the four UDGs plotted. 

Alternatively, these UDGs may exhibit a paucity of dark matter due to an exotic formation pathway. For example, there are extensive arguments in the literature advocating for and against the need to invoke a special formation pathway (e.g., \citealp{vanDokkum2018, Trujillo2019, Silk2019, Emsellem2019, montes2020galaxy-b81, vanDokkum2022, Keim2022, Keim2025}) for the UDG NGC~1052-DF2. A more thorough discussion of this galaxy, and of others that may be similar (e.g., FCC~224; \citealp{Tang2025b, Buzzo2025b}), is beyond the scope of this work. 

It is useful to note that there are also a population of UDGs above the normal relationship in Figure \ref{fig:sigma_mstar}, the vast majority of which have velocity dispersions obtained from their GC system. For non-UDGs, the velocity dispersion of their GC system usually traces the velocity dispersion of their stars. Likewise, for many UDGs, this is the case \citep{Forbes2021}. As such, it is difficult to assess why the distribution of GC velocity dispersions is generally different to that of the stellar velocity dispersions. Those UDGs with elevated GC velocity dispersions are too faint to measure their stellar velocity dispersions. We will therefore consider alternative possibilities. These UDGs are mostly from \citet{Toloba2023}, who studied UDGs in the Virgo Cluster. Being in the cluster environment, it is possible that they are undergoing tidal disruption (as is the case for one UDG being plotted, VLSB-D) and that their velocity dispersion is not indicative of their total mass. Alternatively, being in a cluster environment may make it more likely that intra-cluster GCs are mis-associated with the UDG, artificially driving up their velocity dispersions. The increase of GC-velocity dispersion with local galaxy density shown in figure 11 of \citet{Toloba2023} could be an indicator of either explanation. 

Finally, it is worth noting that the spread in UDG velocity dispersions may be explainable by different dark matter halo profiles. In particular, the elevated velocity dispersions seen for some UDG GC systems could be explained by highly concentrated dark matter halos \citep{Toloba2023}. Alternatively, UDGs with extremely low velocity dispersions may be explainable by extremely low concentration (and/or cored) dark matter halos \citep{TrujilloGomez2022}. A UDG dark matter halo that presents a rising velocity profile with radius may also help to explain the elevated GC velocity dispersions. As the spectroscopically measured GCs tend to trace out to larger radii (in some cases out to $7 R_{\rm e}$), a rising velocity profile with radius would naturally explain GC to stellar velocity dispersions that are above unity. If this were the case, however, we would also expect UDG stellar velocity dispersion to be larger than those of dwarf galaxies of similar stellar mass, due to their being measured in a larger aperture than those of normal dwarfs. 

Despite the length of discussion devoted to the outlying UDGs in Figure \ref{fig:sigma_mstar}, it is worth reiterating the key conclusion, which is that the majority of measured UDG velocity dispersions (particularly UDG stellar velocity dispersions) are normal for their stellar mass. In general, UDGs do not seem to be dynamically hotter or colder than non-UDGs of a similar stellar mass.

\subsection{What drives UDG dark matter content -- Luminosity?}
\begin{figure}
    \centering
    \includegraphics[width=0.45\textwidth]{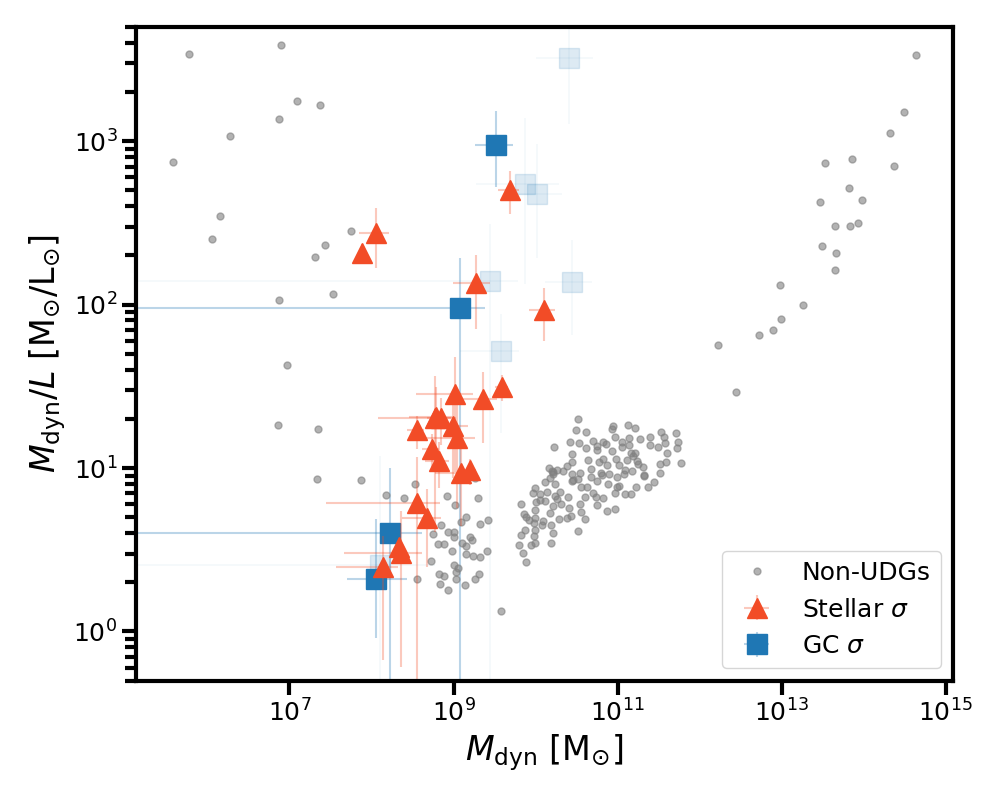}
    \caption{Mass to light ratio vs. dynamical mass within half light radius. The non-UDGs that establish the U-shaped relation are from the works of \citet[][grey points]{Zaritsky2006, Wolf2010, Cappellari2013, Toloba2014} and \citet{Forbes2018}. UDGs are plotted from the catalogue of \citet{Gannon2024b} with dynamical masses coming from their stellar velocity dispersions (red triangles) and their GC system velocity dispersions (blue squares). Where GC velocity dispersions are inferred from $<10$ GCs, we employ a higher transparency to indicate these dispersions may be less reliable.  UDGs exhibit a wide variation in their dark matter properties, with most of them significantly more dark matter dominated than the established U-shaped relation. This is possible evidence of UDGs having halos of total mass more than is expected given the standard stellar mass\,--\,halo mass relationship.}
    \label{fig:mlratio_mdyn}
\end{figure}

In order to best interrogate whether UDGs are dwarf-like or not in their dark matter content, it would be useful to place UDGs on the stellar mass\,--\,halo mass relationship. It is worth pointing out that neither stellar mass, nor halo mass, is a directly observable property. Instead, it is common to use a proxy for both: Luminosity as a proxy for stellar mass; Dynamical mass as a proxy for halo mass. A version of Figure \ref{fig:mlratio_mdyn} has been produced by several authors (e.g., \citealp{Toloba2018, vanDokkum2019b, Gannon2021}). Note that luminosity is only a good proxy for stellar mass if there are reasonably similar stellar populations for the galaxies plotted, and dynamical mass is only a good proxy for total halo masses mass if there are reasonably similar dark matter halo profiles for galaxies being compared. In Figure \ref{fig:mlratio_mdyn} non-UDGs follow a characteristic ``U-shape'' relationship similar to the 
%what is known for the stellar mass\,--\,halo mass relationship when it is plotted as a 
stellar to halo mass ratio vs halo mass relationship (see e.g., figure 2 of \citealp{Wechsler2018}). It is clear, however, that many current UDG dynamical mass measurements do not follow the same U-shape and do not reside on the same locus as non-UDGs of similar luminosity. Many UDGs are much more dark matter dominated within their half-light radius than comparable non-UDGs of similar luminosity (and dwarf-like stellar mass). This presented some of the first evidence that many UDGs may not follow stellar mass\,--\,halo mass relationships that were established without considering their existence.

UDGs, by definition, have larger half-light radii than the majority of galaxies that are of similar luminosity. Naturally, this will lead to larger dynamical masses ($M_{\rm dyn}\propto R_{\rm e} \sigma^2$) as their dynamical mass is being measured within a larger aperture, taking into account more of their dark matter halo and thus increasing the relative dark matter content of the mass measurement \citep{vanDokkum2019b}. This bias cannot, however, rectify the difference between UDGs and non-UDGs seen in Figure \ref{fig:mlratio_mdyn}. Accounting for aperture effects will make many non-UDGs exhibit dynamical mass-to-light ratios more similar to UDGs; however, the great scatter in UDG mass-to-light ratios cannot be accounted for by a simple aperture correction to non-UDG measurements (see e.g., figure 16 of \citealp{vanDokkum2019b} and related discussion). 

\subsection{What drives UDG dark matter content - Globular Clusters?}

\begin{figure}
    \centering
    \includegraphics[width=0.45\textwidth]{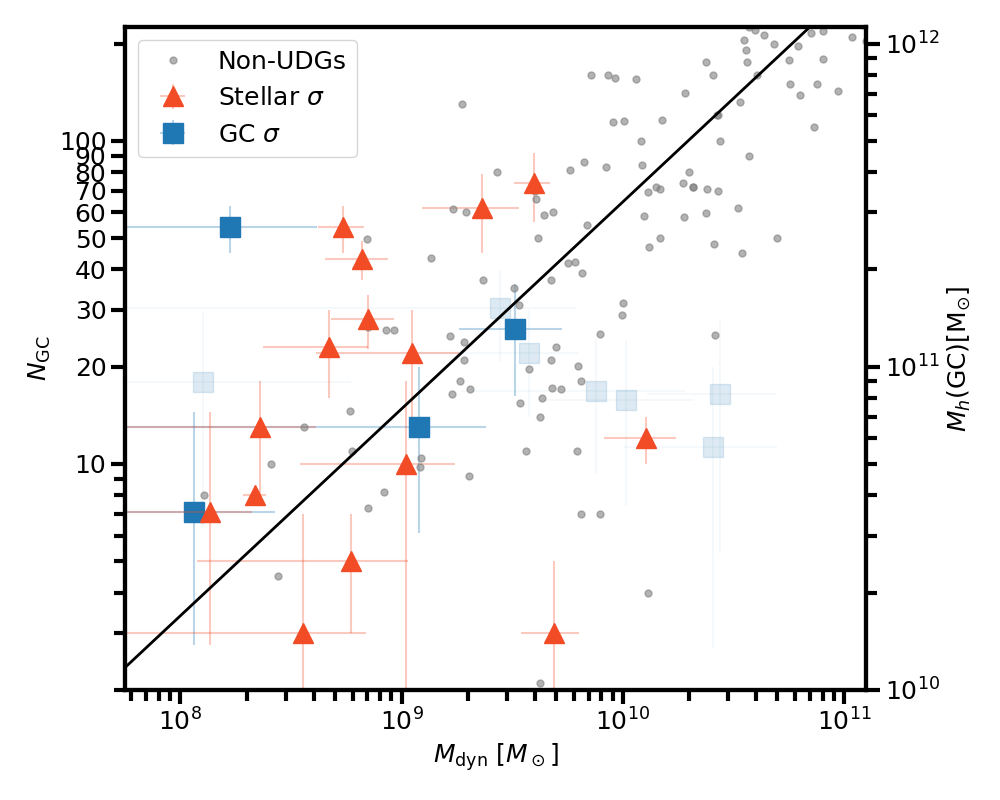}
    \caption{GC system richness ($N_{\rm GC}$) vs. dynamical mass ($M_{\\rm dyn}$). A second y-axis is included where GC-number is translated to a total halo mass estimate using the \citet{Burkert2020} relation. The non-UDGs that establish the relationship are from the catalogue of \citet[][grey points]{Harris2013}. A best-fitting line to these data is included in black. Per previous plots, UDGs are plotted from the catalogue of \citet{Gannon2024b} with dynamical masses coming from their stellar velocity dispersions (red triangles) and their GC system velocity dispersions (blue squares). Per previous plots UDG GC velocity dispersions coming from $<10$ tracers are plotted with a higher transparency. UDGs with well-constrained velocity dispersions largely follow the relationship that has been established for non-UDGs. It has been suggested that this is evidence that they also follow the GC number\,--\,halo mass relationship.}
    \label{fig:ngc_mdyn}
\end{figure}

In Figure \ref{fig:mlratio_mdyn}, many UDGs show extreme dark matter dominance and hints of dark matter halos more massive than what are expected given previously established stellar mass\,--\,halo mass relationships. A separate early sign that they may inhabit massive dark matter halos was their extreme GC richness (e.g., \citealp{Beasley2016, vanDokkum2017}). Given GCs are known indicators of halo mass (e.g., \citealp{2009MNRAS.392L...1S, Harris2013, Burkert2020}), a key question of UDG studies has been to establish if they too follow GC number\,--\,halo mass relationships and thus if GCs may be used to estimate UDG halo masses. We will revisit this question in full in Section \ref{sec:gcnum_halomass}, however, here we will examine the first-order evidence that UDG dynamics correlate with their GC richness. Put another way, if UDG GC counts indicate their total halo masses, then we expect a correlation between UDG GC richness and dynamical mass.

\begin{figure*}
    \centering
    \includegraphics[width=0.95\textwidth]{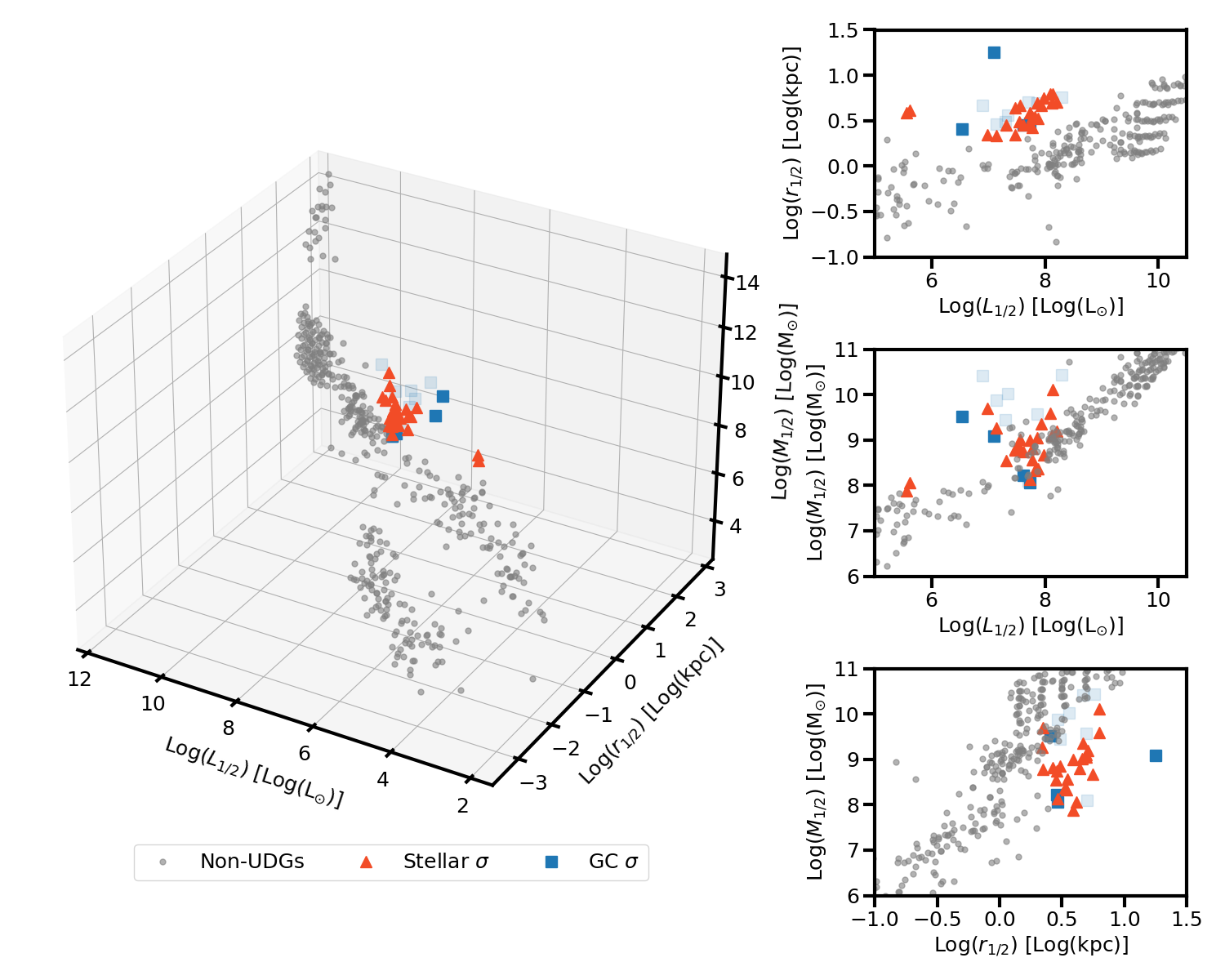}
    \caption{An update on Figure 9 from \citet{Gannon2023}. \textit{Left:} Mass--radius--luminosity space: half-light luminosity ($L_{1/2}$), 3D half-light radius ($r_{1/2}$) and dynamical mass within the half-light radius ($M_{1/2}$). We de-project the plane and zoom around the locus of UDGs on the \textit{Right}. From top to bottom these are the $L_{1/2}$ -- $r_{1/2}$, $L_{1/2}$ -- $M_{1/2}$ and $r_{1/2}$ -- $M_{1/2}$ projections of the plane. We establish the locus for non-UDGs using the data from \citet{Tollerud2011, Toloba2012, mcconnachie2012, Kourkchi2012} and \citet{Forbes2018} (grey points). Per previous plots, UDGs are plotted from the catalogue of \citet{Gannon2024b} with dynamical masses coming from their stellar velocity dispersions (red triangles) and their GC system velocity dispersions (blue squares). Within this parameter space, the primary difference between UDGs and non-UDGs is their larger sizes.}
    \label{fig:fundametnal_plane}
\end{figure*}

In Figure \ref{fig:ngc_mdyn} we plot UDG GC counts vs. their dynamical masses. Non-UDGs that establish the relationship are from \citet{Harris2013}, and a best-fitting line to their relationship is included. In the same way, dynamical mass may be thought of as a proxy for halo mass when creating Figure \ref{fig:mlratio_mdyn}. Here again, we use dynamical mass as a proxy for halo mass. It is worth noting that this assumption strictly assumes a level of self-similarity of UDG dark matter halo profiles whereby any correlation may be diluted, were some UDGs to exhibit strong cores, while others exhibit strong cusps, at the same total halo mass. It is evident from Figure \ref{fig:ngc_mdyn} that UDGs, particularly those with dynamical masses measured from their stellar velocity dispersions, follow the relationship that has been established for non-UDGs. We take their following the relationship as first-order evidence that UDGs indeed follow the GC number\,--\,halo mass relationship and that many UDGs reside in massive dark matter halos. Further, it should be noted that the relationship may tighten further when considering the stellar mass contained within the GC system. Some UDGs exhibit low numbers of GCs, but a top-heavy mass function (e.g., \citealp{Janssens2022}), which may skew them off the relationship to the right in Figure \ref{fig:ngc_mdyn}.

Our evidence that UDGs may indeed follow the GC number\,--\,halo mass relationship has two important corollaries for their formation. Firstly, if UDG total halo masses can be estimated via their GC counts, their dark matter halo profile shape can be inferred from individual dynamical mass measurements. This has been done by various authors (e.g., \citealp{Beasley2016, Toloba2018,  Gannon2020, Gannon2022, Forbes2021, Levitskiy2025}) who generally find cored (or equivalently low concentration) dark matter halos best recover the measured dynamical masses of (usually GC-rich) UDGs. UDGs residing within cored or possibly low-concentration dark matter halos may agree with their formation via strong stellar feedback \citep{Toloba2018} or may help with their formation via tidal effects \citep{Carleton2019}. As a counterpoint to the above evidence, the study of \citet{Kravtsov2024} built a toy model to explain UDG dynamical masses and found the opposite - that many of the most massive UDGs may have higher, not lower, concentration dark matter halos. 
Secondly, and simply, if UDG GC content indicates that many UDGs reside in massive dark matter halos, the most GC-rich UDGs cannot simply be an extension of the regular dwarf galaxy population, as they reside in dark matter halos that are an order of magnitude more massive. 

\subsection{Where are UDGs on the Fundamental plane of Elliptical Galaxies?}
In Figure \ref{fig:everything_plot}, we established UDGs as an extension of (primarily spheroidal) stellar systems to a new region of previously unexplored parameter space, and so it is of interest where UDGs reside on the Fundamental plane of elliptical galaxies \citep{Djorgovski1987}, which tend to be pressure supported. Rather than plotting the plane in its traditional effective radius\,--\,velocity dispersion\,--\,surface brightness form, we place UDGs in an altered form of the plane comprising dynamical mass\,--\,effective radius\,--\,luminosity in Figure \ref{fig:fundametnal_plane}, as it represents a simple extension of Figure \ref{fig:everything_plot} in the direction of dynamical mass. 

It is clear from Figure \ref{fig:fundametnal_plane} that the largest difference between UDGs and other elliptical systems is simply their increased half-light radii. This is particularly apparent in the half-light radius\,--\,luminosity and dynamical mass\,--\,half-light radius de-projections of the plane. At face value the fact that UDGs are not outliers in the dynamical mass\,--\,luminosity projection of the plane may seem at odds with previous conclusions from Figures \ref{fig:mlratio_mdyn} and \ref{fig:ngc_mdyn}. Interestingly,  \citet{Kadowaki2021} show in their appendix A how the large half-light radius of UDGs, and the current bias towards spectroscopy of the largest UDGs, creates a bias towards UDGs residing in massive dark matter halos.

Furthermore, \citet{Zaritsky2023c} and \citet{Zaritsky2023} show how a fundamental plane type argument can be used to estimate UDG halo masses from their photometric properties. Again, UDGs are found to reside in dark matter halos of greater mass than similar stellar mass dwarfs, with a primary driver of this conclusion being their large half-light radii. A key conclusion of their work is that UDGs may present at least two times and frequently greater than ten times less efficient star formation than dark matter halos of a similar mass that do not host UDGs \citep{Zaritsky2023}. This conclusion would support the scenario where some UDGs are ``failed galaxies" that have failed to form the stellar mass expected given their massive dark matter halos. 

Indeed, much of the evidence currently gathered from UDG dynamics, their elevated mass to light ratios (Figure \ref{fig:mlratio_mdyn}), their following the GC number\,--\,dynamical mass relationship (Figure \ref{fig:ngc_mdyn}) and their fundamental plane positioning (Fig. \ref{fig:fundametnal_plane}), all support the hypothesis that a significant fraction of the UDGs studied for dynamics may follow a ``failed galaxy'' formation pathway. We stress the importance of the caveat that the conclusion can only be reached for the (biased) sample of UDGs for which we currently have high-quality spectroscopy that enables dynamical measurements. As this sample has been shown to be generally larger and more GC-rich than the general UDG population \citep{Gannon2024b}, it is likely also biased to the ``failed galaxy'' formation pathway, in comparison to the general UDG population. 

%%%%%%%%%%%%---- STELLAR POPS
\section{UDG Internal Properties II: Clues from stellar populations}\label{sec:stellar_pops} 
The study of the stellar populations of galaxies is key to understanding their formation processes. It can provide a multitude of details, e.g., the age of the stars forming the galaxy, how they formed over time (e.g. long and steady, bursty, or very quick and early star formation histories), and what was their chemical enrichment by means of the total metallicity and the relative abundances of certain elements (e.g., [Mg/Fe]). As detailed in Figure\,\ref{fig:glance_table}, these properties can help differentiate between the diverse UDG formation scenarios. Only a couple of local UDGs have been studied via resolving individual stars within the galaxy (e.g. WLM by \citealt{Albers2019}, And\,XIX by \citealt{Collins2022}, and F8D1 by \citealt{Smercina2025}). Thus, the vast majority of the stellar population information to date comes from integrated measurements of the stellar populations of UDGs. 

\subsection{Global stellar populations - are UDGs the continuation of classical dwarfs?}
Given the possible formation scenarios discussed in Sect.\ref{sec:formation_scenarios}, the obvious question is whether UDGs have stellar population properties that are more similar to classical dwarfs or whether they resemble some other types of galaxies. The first constraints on the stellar populations of Coma UDGs came from broad-band colours \citep{vanDokkum2015}, which suggested either old, metal-poor populations or somewhat younger, more metal-rich ones. Similar results were obtained by \citet{vanderBurg2016}, \citet{Roman2017}, and \citet{Prole2018} for UDGs in other environments, showing some environmental differences: cluster UDGs tended to be red and quenched, while field UDGs were often bluer and still forming stars. This supported the idea that star-forming UDGs may fall into clusters, quench, and evolve into the red systems observed today (e.g. \citealt{Roman2017b,Prole2019b,Roman2019})

However, given the age\,--\,metallicity degeneracy resulting from the colours alone, spectroscopy was needed to obtain more constrained results. Over the last decade a huge effort has been carried out to obtain spectra to study the stellar populations of UDGs via full-spectral-fitting or classical line indices. These studies revealed a large variety in the stellar population properties of UDGs \citep{vanDokkum2015b,Kadowaki2017,Gu2018, FerreMateu2018, RuizLara2018, Chilingarian2019,FerreMateu2023}. While many of the UDGs are indeed old ($\sim7-9$\,Gyr) and metal poor ([$Z$/H]$\sim -$1\,dex), dominated by elevated [Mg/Fe] values ($\ge$0.4\,dex), and showing very early and fast star formation histories (SFHs), others are found to be much younger (3--5\,Gyr), presenting extended SFHs, and with metallicities more similar to classical dwarfs ([$Z$/H]$\sim -$0.5\,dex).  It is worth mentioning that these works were focused on the cluster environment, and only a few isolated or low density environment UDGs have spectroscopic results \citep{Roman2019, MartinNavarro2019, FerreMateu2023}, with results comparable to those in clusters. SED fitting studies have thus taken over to measure not only large, complete UDG samples, but also significantly stepping away from the cluster environment (e.g. the SMUDGes survey \citealt{Zaritsky2019, Barbosa2020}, or the works of \citealt{Buzzo2022, Buzzo2024, Buzzo2025}, which combined the results from the MATLAS survey and other environments). These surveys now provide stellar population estimates for hundreds of UDGs, allowing comparisons across environments and with simulations. 

\begin{figure}
    \centering
    \includegraphics[width=0.99\linewidth]{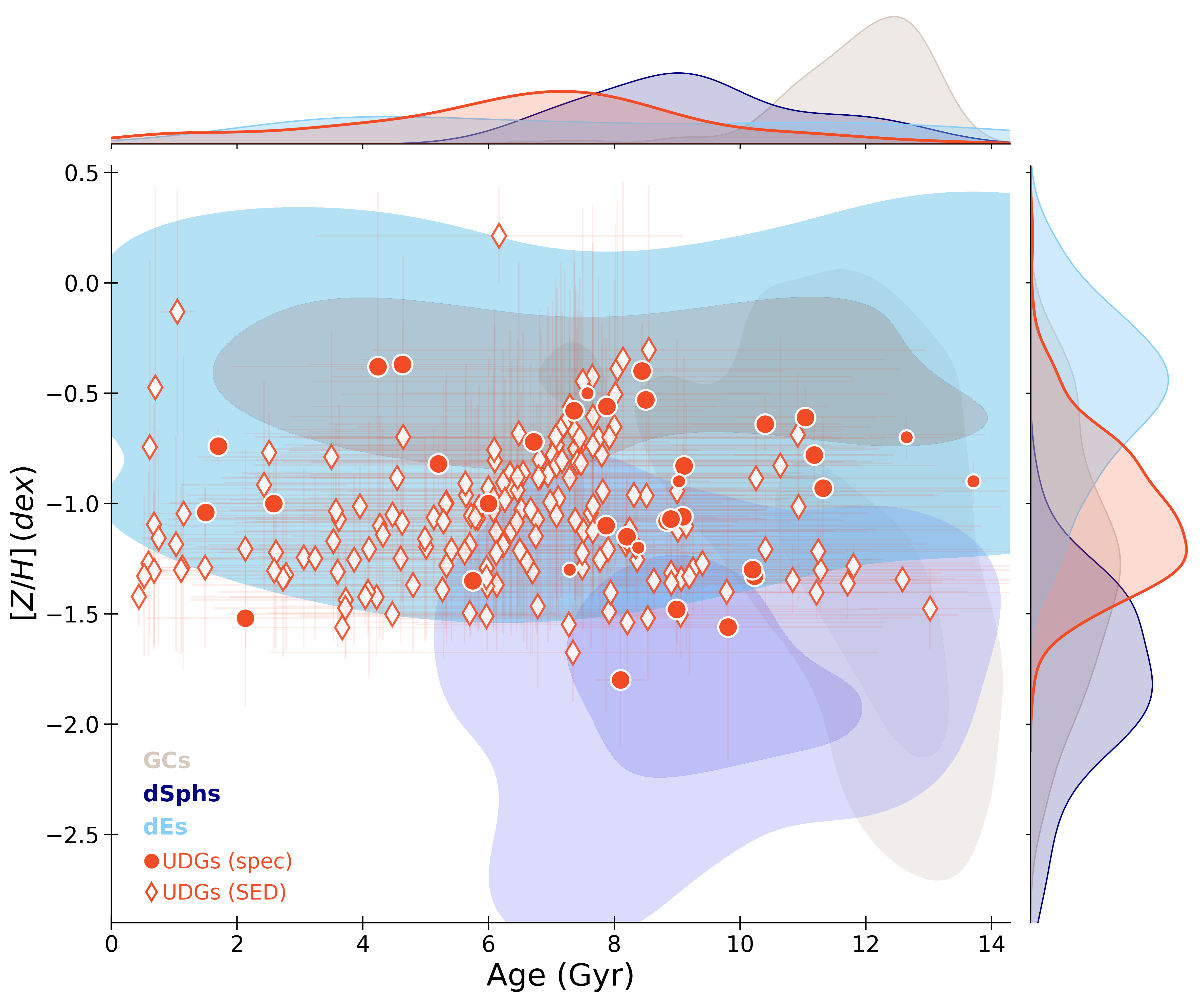}
    \caption{Age\,--\,metallicity relation containing the spectroscopic samples (open circles) from \citet{FerreMateu2023} with updates on their figure to include the results of \citet{Levitskiy2025, Buzzo2025b} and \citet{Doll2025}. Photometric results for the SMUDGEs galaxies \citep{Barbosa2020} and those presented in \citet{Buzzo2022, Buzzo2024} are also included, shown by open diamonds, creating a sample of 212 UDGs in total. The figure also shows the distributions in age and metallicity of different types of galaxies for comparison, as specified in the legend, from: \citet{Janz2016}, \citet{FerreMateu2018}, \citet{Recio-Blanco2018}, \citet{Naidu2022}, and \citet{RomeroGomez2023}. Overall, UDGs span the entire age range, from very young to very old ones (1 to 14\,Gyr), but are more limited in total metallicities to [$Z$/H]$>-1.5$\,dex. It can be seen that overall UDGs are compatible with being the envelope of the bulk of the age\,--\,metallicity distribution of dEs, but particularly for the older ones the locus is also shared with the GCs and dSphs distributions.}
    \label{fig:AgeZ_all}
\end{figure}

The age\,--\,metallicity distribution of the compilation of 212 observed UDGs is shown Figure \ref{fig:AgeZ_all}. The figure is updated from \citet{FerreMateu2023} to incorporate all the UDGs currently with spectroscopic measurements discussed above (filled circles), and also those with SED fitting stellar population results (open diamonds). The density contours and distributions correspond to the stellar populations of comparison galaxies such as dEs (light blue), dSphs ({dark blue) and GCs (light brown}; see \citet{FerreMateu2023} for a full description of these comparison samples). Overall, the bulk of red, quenched UDGs span all possible ages, with a median value of $\langle$Age$\rangle=7.1\pm$2.7\,Gyr (see also Table \ref{tab:1}), but are more limited in total metallicities, with an average of $\langle$[$Z$/H]$\rangle=-1.08\pm$0.34~dex. UDGs overall seem to occupy the lower envelope of the dE distribution. Except for the younger UDGs, which mostly match dEs, the age\,--\,metallicity relation is not enough to decipher the origin of older UDGs, which show metallicities that resemble either dEs, dSphs and GCs. In particular, the oldest UDGs could also track the GCs distribution. 

\subsubsection{The Mass\,--\,Metallicity relation of UDGs}
\begin{figure}
    \centering
    \includegraphics[width=0.99\linewidth]{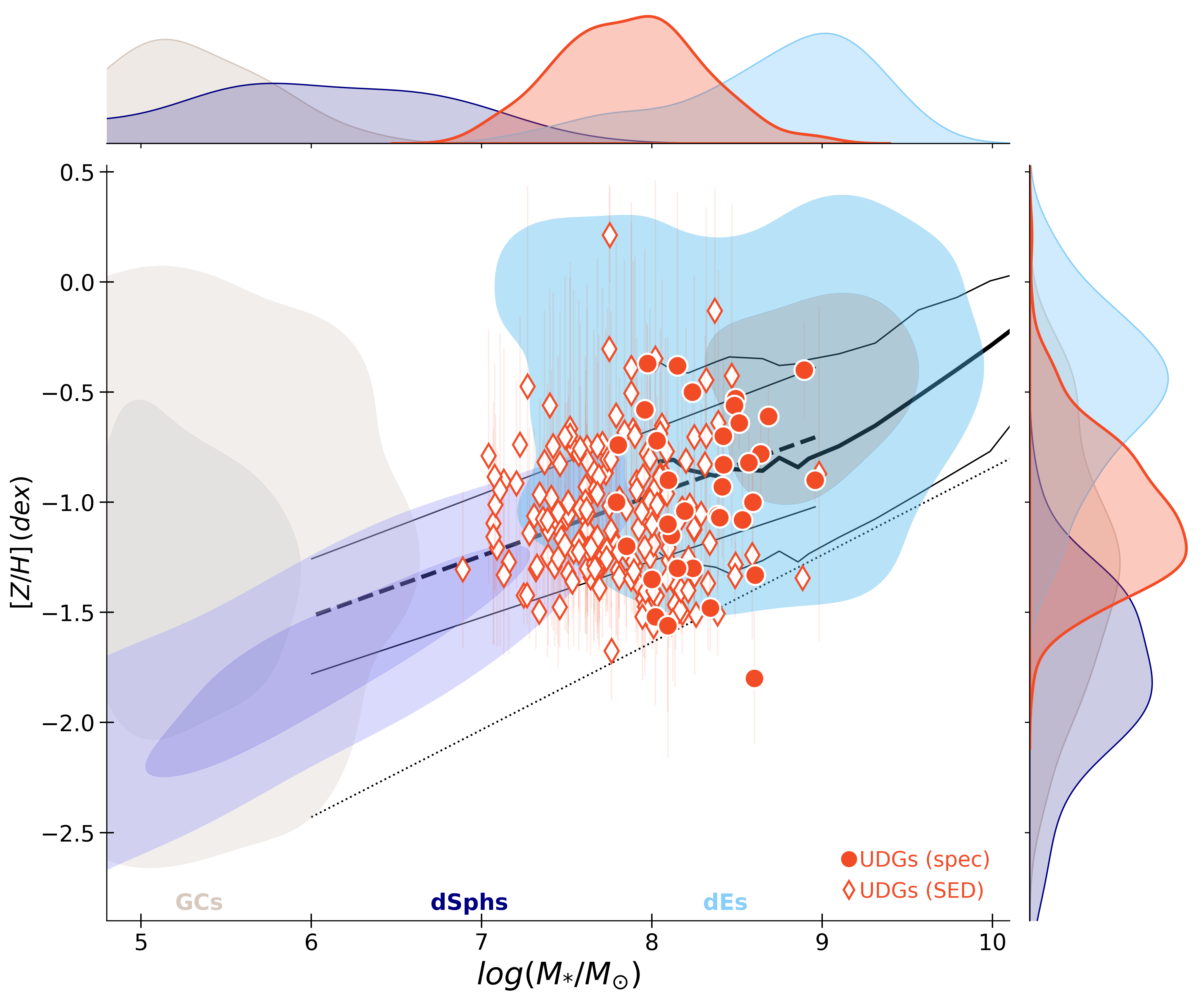}
    \caption{The figure shows an updated version of the stellar mass\,--\,metallicity relation (MZR) from \citet{FerreMateu2023}. The MZRs and their intrinsic scatter of local massive and low-mass galaxies (dashed line for \citealt{Panter2008} and solid line for \citealt{Simon2019}, respectively) are shown, as well as the theoretical prediction of $z\sim 2$ galaxies (dotted line, \citealt{Ma2016}). Similar to Figure\,\ref{fig:AgeZ_all}, it shows the spectroscopic and SED fitting sample of UDGs, and the distribution of dEs, dSphs, and GCs in coloured contours. It can be seen that the bulk of the UDGs scatter around the MZRs, with similar metallicity to classical dwarfs at their stellar mass. However, several UDGs inhabit the parameter space around the $z\sim 2$ relation, possibly indicative of a `failed-galaxy' origin. }
    \label{fig:mzr_all}
\end{figure}

One of the key scaling relations in stellar population studies is that of the stellar mass\,--\,metallicity relation (MZR, e.g. \citealt{Gallazzi2005, Panter2008, Kirby2013}). Figure \ref{fig:mzr_all} shows the same UDGs and comparison galaxies as in Figure \ref{fig:AgeZ_all}, together with the local MZR of both low- and high-mass galaxies (\citealt{Kirby2013} and \citealt{Panter2008}, respectively), and the MZR of $z\sim$2 galaxies from \citet{Ma2020}. The figure shows that UDGs fill the gap between dEs and dSphs in terms of stellar mass, and overall scatter around the scaling relations at a given stellar mass, consistent with a continuation of the dwarf sequence (e.g. puffed-up dwarfs). Yet, some outliers seem to exist. Some UDGs exhibit higher-than-expected metallicities, which might be interpreted as remnants of tidally stripped more massive galaxies (e.g. \citealt{Liao2019}; \citealt{Sales2020}; \citealt{Tremmel2020}; \citealt{Zheng2025}). Conversely, there is a non-negligible fraction (roughly 10\% of the entire sample) of galaxies with lower-than-expected metallicities that instead follow the $z\sim 2$ MZR. Interestingly, the majority of them are GC-rich (see \citealt{Buzzo2025}) and show the most elevated [Mg/Fe] (see \citealt{FerreMateu2023} and next section), thus suggesting a formation as "failed galaxies'' that quenched early before significant metal enrichment (\citealt{FerreMateu2023}, \citealt{Buzzo2025}). 

An alternative way to elucidate the formation pathways of UDGs is to compute the deviation from the MZR, $\delta_{\rm{MZR}}$ (e.g. \citealt{Barbosa2020, Buzzo2024, Buzzo2025}) and study whether there is any dependence of this deviation on other structural and stellar population parameters. \citet{Buzzo2025} found that UDGs that aligned with the traditional dwarf MZR tend to be younger, exhibit extended SFHs, contain relatively few GCs, and have lower GC-to-stellar mass ratios. In contrast, UDGs falling below the dwarf MZR are generally older, have early and fast SFHs, and are rich in GCs. Notably, UDGs with the highest GC system masses relative to their stellar mass are found below the local MZR, inhabiting a higher redshift MZR locus as may be expected if they are `failed galaxies'. We will discuss this further in Section \ref{sec:twoclasses}.

Finally, we wish to include a note of caution to those assembling their own MZR relations for UDGs. Although the MZR is one of the most powerful relations to understand the nature of galaxies, it also presents some intrinsic caveats. For instance, some works present [Fe/H] as total metallicity, which is not necessarily the case for low-metallicity galaxies, as the low-metallicity stars used to compute the stellar population synthesis models are $\alpha$-enhanced. There is the issue of the inhomogeneity and incompleteness of the different samples (see e.g. figure 6 from \citealt{Buzzo2024}). Lastly, different techniques can introduce systematics in the recovery of certain stellar population properties (see e.g. \citealt{Buzzo2022, Webb2022, FerreMateu2023, Buzzo2024}. One avenue to explore in the future would be to combine them in spectrophotometric studies, which seem to render the most constrained results (see e.g. \citealt{Webb2022,Tang2025a, Tang2025b}).

\subsubsection{The elevated [Mg/Fe] values of UDGs}
Another key property for understanding UDG formation is the relative abundance of $\alpha$-elements such as Mg compared to Fe. These elements are produced by supernovae species that operate on different timescales, making them a useful cosmic clock of star formation. High [Mg/Fe] values indicate rapid, early star formation that ended before type Ia supernovae contributed significant Fe, while lower values imply more extended star formation and greater Fe enrichment (e.g. \citealt{Thomas2005}; \citealt{delaRosa2011}; \citealt{McDermid2015}).

\begin{figure}
    \centering
    \includegraphics[width=0.99\linewidth]{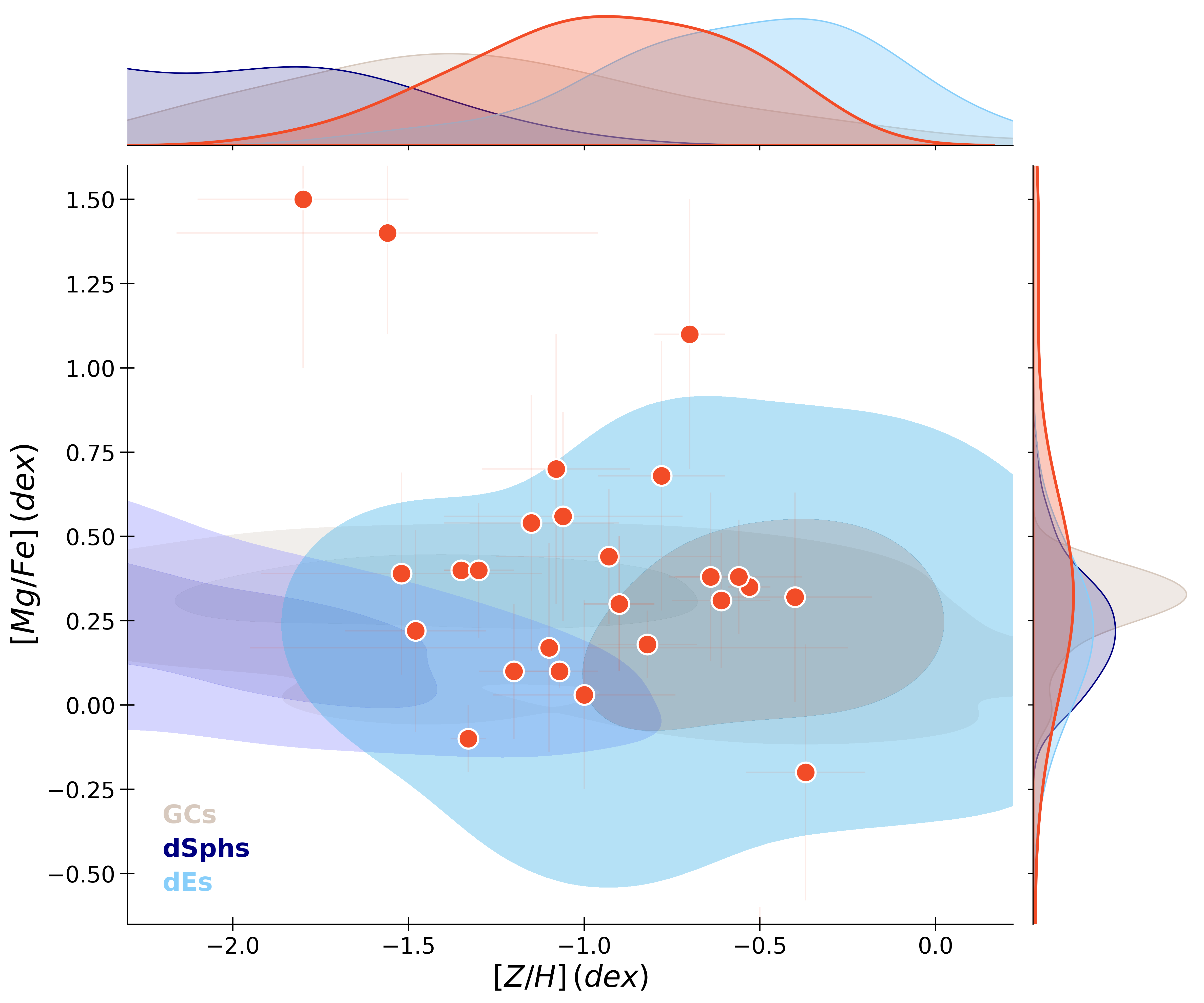}
    \caption{Similar to Figs. \ref{fig:AgeZ_all} and \ref{fig:mzr_all}, now representing the [Mg/Fe]--metallicity relation. Only UDGs studied with high S/N spectroscopy are shown, as SED fitting cannot provide this property. While the majority of UDGs show $\alpha$-enhancement values compatible with dEs, GCs or dSphs, there are three clear outliers with [Mg/Fe]$>1$\,dex that inhabit a region of parameter space that was previously empty of galaxies.}
    \label{fig:alfZ_all}
\end{figure}

Most UDGs studied so far show elevated $\alpha$-enhancements (typically [Mg/Fe]$\sim$0.4\,dex; \citealt{FerreMateu2023}). In the latter, it was shown that a correlation between the [Mg/Fe] and the formation timescales did exist and that indeed those with the highest elemental abundances where the ones that formed and quenched the earliest. We cation that this could be possibly biased by the fact that the majority of these UDGs are GC-rich and in clusters. We show the [Mg/Fe]--[$Z$/H] relation in Figure \ref{fig:alfZ_all}, which shows that the majority of UDGs follow the locus of dEs, but also GCs and some dSphs. 

However, three objects clearly stand out in the figure, presenting extremely high values of [Mg/Fe]$\sim$1--1.5\,dex, values not even seen in the most metal-poor stars of the Milky Way halo. These puzzling cases (DGSAT\,I, \citealt{MartinNavarro2019}; Y358, \citealt{FerreMateu2023}; and Hydra-I UDG9, \citealt{Doll2025}) do not fit easily within conventional enrichment scenarios and may reflect unusual processes such as Fe depletion rather than Mg enhancement,  or an extremely early, truncated star formation, i.e. `failed galaxies'. 

\subsection{What is the impact of local (global) environment on the stellar populations of UDGs?}
At this point, it is clearly established that UDGs are a mixed-bag of objects, representing everything from being puffed up dwarf galaxies, to failed galaxies, and even possibly tidally stripped remnants of more massive galaxies. The diversity underscores that UDGs are not a homogeneous class, but rather a mixture of systems whose histories are imprinted in their ages, metallicities, and abundance ratios, amongst other key structural properties. The next logical question is to ask whether environment (either global or local) has any say in shaping the nature of UDGs. 

\begin{figure*}
    \centering
    \includegraphics[width=0.95\linewidth]{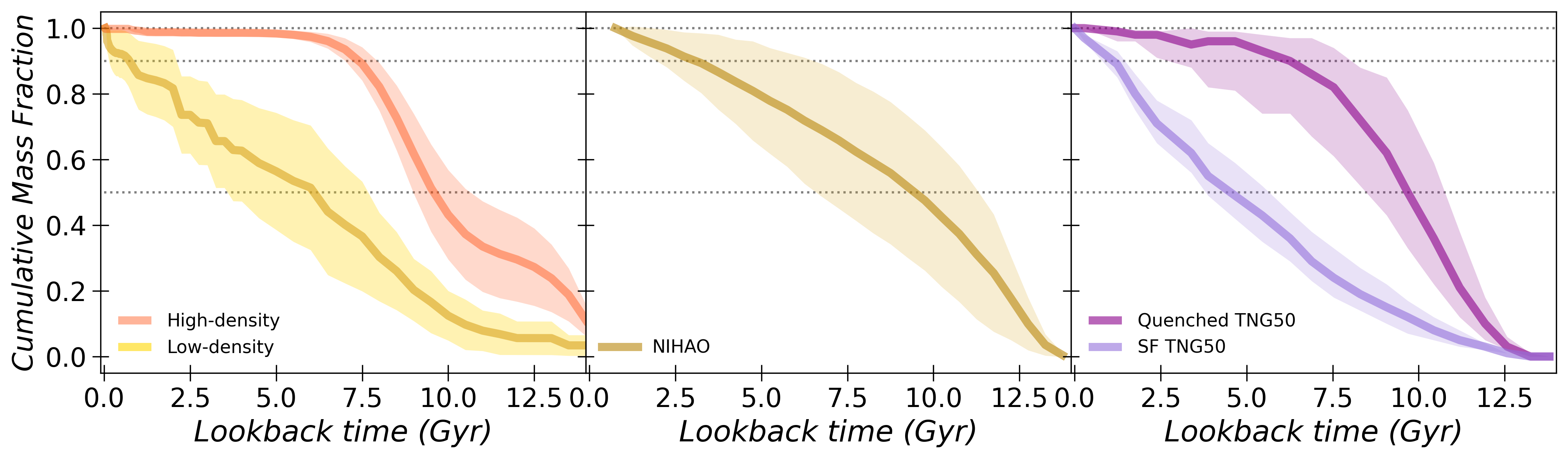}
    \caption{Star formation histories adapted from \citet{FerreMateu2023} (left, observations), \citet{CardonaBarrero2022} (middle, NIHAO, priv. comm), and \citet{Benavides2024} (right, TNG50). Each panel shows the cumulative mass fraction over cosmic time, coloured by the local environment: yellow-ish tones for low-density UDGs and orange for high-density ones (left panel for observations and middle one for NIHAO); or by star formation status: quenched in violet and star forming in purple (right panel, TNG50 UDGs). The dotted lines mark 50 and 90\% of mass fraction assembled (used to compute the lookback times $t_{50}$ and $t_{90}$). While quenched UDGs in TNG50, high-density observed UDGs, and NIHAO UDGs all show similar $t_{50}$ ($\sim9.5$\,Gyr), UDGs in low-density environments show much longer timescales of formation, with $t_{50}\sim6.0$\,Gyr. However, both NIHAO and low-density UDGs show the longest quenching timescales (as seen by the lookback time they formed 90\%, $t_{90}\sim 1-3$\,Gyr), while quenched TNG50 and high-density UDGs have similarly faster ones ($t_{90}\sim 7$\,Gyr), quenching in only a couple of Gyr after half of their stellar mass was built.
    }
    \label{fig:SFHs_all}
\end{figure*}

UDGs in galaxy clusters have become a central focus in the past decade, particularly regarding their stellar populations. Understanding the role of a galaxy’s location within the cluster is therefore essential. \citet{FerreMateu2018} showed that, although Coma UDGs' stellar populations do not exhibit a strong cluster-centric dependence overall, those in the core appeared more metal-rich than systems in the outskirts previously studied by \citet{Kadowaki2017} and \citet{Gu2018}. However, such trends may be affected by projection effects, which has motivated the widespread use of phase-space diagrams to examine correlations between cluster location and UDG properties (e.g. \citealt{Alabi2018, Gannon2022, FerreMateu2023, Forbes2023}). These diagrams plot galaxy velocity relative to the cluster mean (normalised by the velocity dispersion) against projected cluster-centric distance (normalised by the virial radius). Different regions of this phase-space correspond to different infall epochs \citep{2017ApJ...843..128R}. While informative, this diagnostic is not without limitations: statistical contamination from interlopers may assign galaxies to incorrect infall regions. As such, trends inferred from these diagrams should be interpreted with caution.

\begin{table}
\centering
\caption[]{Average stellar population properties of UDGs from spectroscopic and photometric studies (\dag \,only spectroscopic). }
\begin{tabular}{c c c c }   
                    & \textbf{All}   &  \textbf{High-density} & \textbf{Low-density}  \\ 
\hline
%Number               &    212                 &  43                  & 168            \\
\#spec/\#SED           &    35/177            &  29/14             & 6/163        \\
\hline
Age (Gyr)            &    7.0$\pm$2.7         &  8.4$\pm$2.8         & 6.5$\pm$2.5    \\
$[$Z/H$]$ (dex)      &    $-$1.08$\pm$0.31    &  $-$1.01$\pm$0.34    & $-$1.10$\pm$0.31 \\
$[$Mg/Fe$]$(dex)\dag &    0.37$\pm$0.45       &  0.38$\pm$0.43       & 0.54$\pm$0.55  \\
t$_{50}$\,(Gyr)\dag  &    8.9$^{+0.9}_{-1.0}$ &  9.5$^{+1.0}_{-0.6}$ & 6.2$^{+1.4}_{-2.6}$ \\
t$_{90}$\,(Gyr)\dag  &    2.7$^{+4.7}_{-2.0}$ &  7.4$^{+0.5}_{-0.5}$ & 0.7$^{+1.3}_{-0.6}$ \\
\hline
\end{tabular}
\label{tab:1}
\end{table}

If galaxies typically quench $\sim$1.5\,Gyr after entering the cluster environment \citep{Muzzin2014}, those quenched earliest should be expected to be concentrated in the cluster centre (i.e. in the `very early infall' region). Observations, however, reveal a more complex picture. \citet{FerreMateu2023} and \citet{Doll2025} report a broad range of quenching times—defined as the epoch when the galaxy has formed 90\% of its stellar mass ($t_{90}$). This finding may be complicated by the fact that some of the UDGs could be quenched prior to infall. Overall, no strong correlations have been identified between infall time, position in the phase-space, and structural properties such as size or stellar mass. In terms of stellar populations, [Mg/Fe] shows the clearest trend: the most $\alpha$-enhanced UDGs are found near the cluster centre and exhibit evidence for rapid quenching \citep{FerreMateu2023}, consistent with results for non-UDGs of similar mass (e.g. \citealt{Pasquali2019, Smith2009, Gallazzi2021}). By contrast, stellar age and metallicity show little dependence on infall region, in line with trends seen in other dwarf galaxies (e.g. \citealt{RomeroGomez2023}). 

All in all, we note that while the phase-space can be a useful tool to understand the formation pathways of individual UDGs, it still needs to be aided by other indicators. Moreover, while the LEWIS sample (see \citealt{Iodice2020b, Iodice2023}) is the most homogeneous in terms of cluster members, the number of the observed UDGs in that cluster is still limited to fewer than ten and covers only out to 0.4 virial radii, thus one of the next crucial steps will be to obtain larger, more complete samples of UDGs within a given cluster.

Environmental comparisons reveal clear differences in the stellar populations of UDGs. Systems in clusters are generally older, while those in the field tend to be younger (\citealt{Barbosa2020,Buzzo2022}). Metallicity also varies with environment: field UDGs are typically more metal-poor but follow the MZR more closely \citep{Buzzo2024}, whereas cluster UDGs are slightly more metal-rich yet display a larger scatter around the relation (\citealt{FerreMateu2023, Doll2025}). These contrasts align with their star formation histories (SFHs), which are usually more extended in field UDGs but shorter or quenched earlier in cluster systems (see e.g. \citealt{FerreMateu2018, RuizLara2018, MartinNavarro2019, Villaume2022, FerreMateu2023, Levitskiy2025}). 

The influence of the local environment appears to be especially important. Even within clusters, some UDGs reside in filaments or are on their first infall, conditions that are more akin to groups or the field than high density environments as galaxy clusters. UDGs in high-density regions seem to be both older (8.4 vs. 7.7\,Gyr) and more metal-rich ($-1.01$ vs. $-1.16$\,dex) than their low-density counterparts, as shown in Table \ref{tab:1}.

Star formation histories further highlight these distinctions \citep{FerreMateu2023}. Left panel of Figure \ref{fig:SFHs_all} shows that UDGs in dense environments typically show rapid formation and early quenching, with lookback times of $t_{50}=9.5^{+1.0}_{-0.6}$Gyr and $t_{90}=7.4^{+0.5}_{-0.5}$Gyr (being $t_{50}$ and $t_{90}$ the time it took to build 50 and 90\% of the galaxy's stellar mass, respectively). Those in low-density environments, by contrast, often sustain star formation for longer, quenching much later or in some cases, have not yet quenched, with $t_{50}=6.2^{+1.4}_{-2.6}$Gyr and $t_{90}=0.7^{+1.3}_{-0.6}$Gyr. The most elevated [Mg/Fe] values are overall compatible with the early and fast formation of high-density galaxies (see figure 4 of \citealt{FerreMateu2023}). We note, moreover, that delayed SFHs as those in low-density environments of Figure\,\ref{fig:SFHs_all} are compatible with those of low-mass local dwarfs (e.g. \citealt{Weisz2014a, Weisz2019}), or similar to the recovered SFHs of the Local Volume UDGs from resolved stars (e.g. \citealt{Albers2019, Collins2022, Smercina2025})

\subsection{Do simulated UDGs have similar stellar populations to observed ones?}
As discussed in Sect. \ref{sec:formation_scenarios}, the formation of UDGs has been explored through a variety of simulation frameworks, each highlighting different physical mechanisms, although there is a clear bias towards the `puffed up' origin.  Notably, while many simulations successfully produce UDGs, they often fail to generate some aspects of more compact classical dwarf ellipticals within the same stellar mass range--a discrepancy that underscores the `dwarf diversity' challenge in galaxy formation models \citep{Sales2022}.

In low-density environments, internal processes such as stellar feedback are expected to play a key role. The NIHAO simulations \citep{DiCintio2017, Jiang2019, CardonaBarrero2020} show that repeated energy injection from supernovae induces potential fluctuations that expand the stellar component, producing galaxies with extended, bursty SFHs, low metallicities, and large sizes. The middle panel of Figure \ref{fig:SFHs_all} shows the averaged SFHs of NIHAO UDGs (Cardona-Barrero, priv. comm.). While these UDGs form relatively quickly, they quench later than most observed systems in low-density environments.

\begin{figure*}
    \centering
    \includegraphics[width=0.95\linewidth]{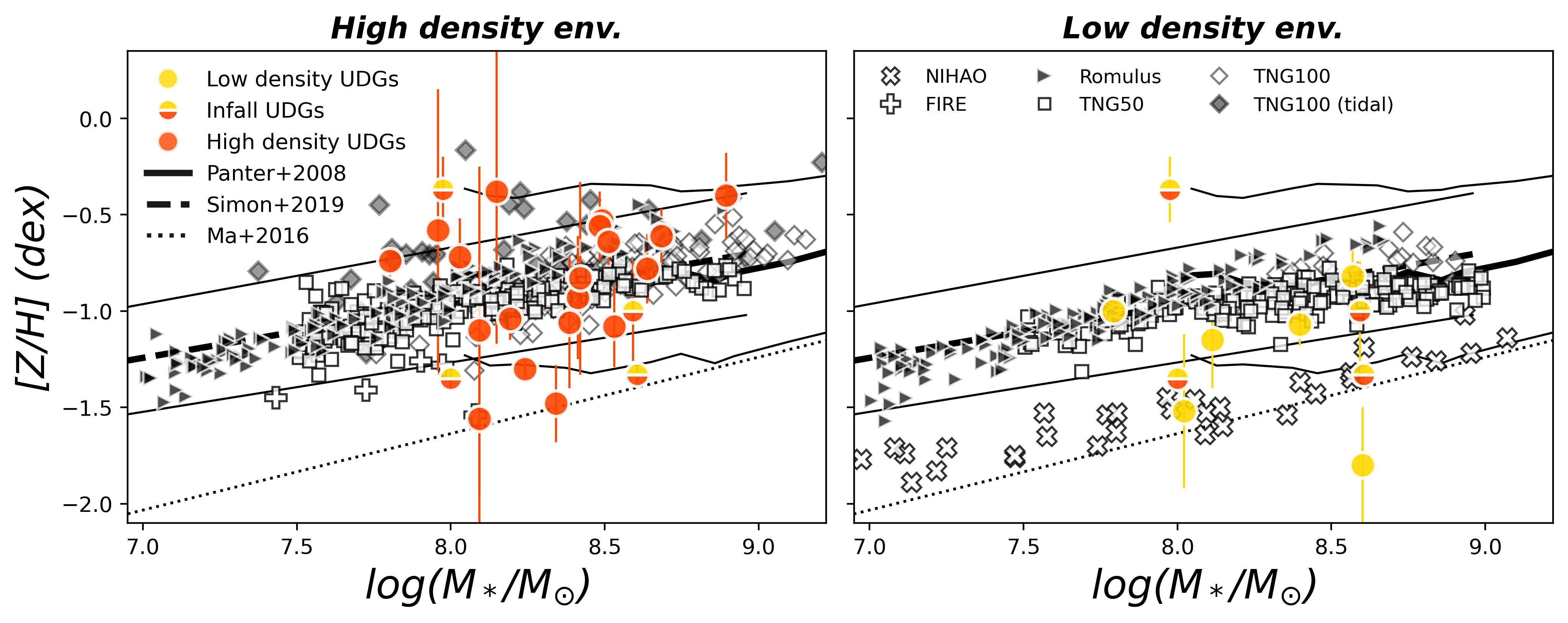}
    \caption{Figure adapted from \citet{FerreMateu2023} to show the MZR as in Figure \ref{fig:mzr_all}, for both observed (coloured symbols) and simulated UDGs (white, grey, and black symbols), now divided by local environment: high-density galaxies (cluster, left panel) and low-density ones (field, group, and infall/filaments, right panel). The local scaling relations at different stellar masses are shown in thick solid and dashed lines (\citealt{Panter2008} and \citealt{Simon2019}, respectively), together with their intrinsic scatter (thin solid lines). The theoretical relation for $z\sim$2 galaxies \citep{Ma2016} in shown as a dotted line. Per the legends, UDGs considered to be infalling are represented in both panels in a split coloured symbol. FIRE (white crosses; \citealt{Chan2018}), RomulusC (black triangles; \citealt{Tremmel2020}), TNG100 (white diamonds for cluster UDGs, shaded grey for those identified as tidally stripped galaxies; \citealt{Sales2020}), and TNG50 (white squares; \citealt{Benavides2024}) are depicted in the high-density panel. Romulus25 \citep{Wright2021}, TNG50, and TNG100 also appear in the low-density category, alongside UDGs with NIHAO (white x-symbols; \citealt{DiCintio2017}). While simulations produce UDGs with a relatively tight scatter in their MZR, observed UDGs are found to have a much larger scatter. In particular there is a population of low metallicity UDGs that do not appear to be well reproduced by any of the simulations, i.e. those of the `failed galaxy' type.}
    \label{fig:mzr_sims}
\end{figure*}

Environmental effects further shape UDG properties, particularly in clusters. The  RomulusC (cluster) and Romulus25 (field) simulations \citep{Tremmel2017, Tremmel2020, Wright2021} suggest that early major mergers can increase angular momentum and expand the stellar distribution. In cluster environments, early infall enhances passive evolution, while in the field, merger-driven puffing-up dominates.  These models align well with observational signatures such as low central surface brightness and intermediate stellar ages. Similarly, FIRE simulations \citep{Chan2018} produce cluster UDGs through internal feedback, though quenching is artificially imposed to match cluster infall times. These models reproduce realistic sizes and morphologies, but the artificial quenching poses a challenge to directly compare to observations. 

Illustris-TNG simulations (TNG50, \citealt{Benavides2022, Benavides2024} and TNG100, \citealt{Sales2020}) extend this picture by distinguishing between `born' UDGs (formed as diffuse systems before infall) and `tidal' UDGs (initially more compact galaxies that were expanded through cluster tides and interactions). This framework reproduces the diversity of cluster UDGs, with cluster members tending to be smaller, redder, and more metal-poor, while systems in low-density environments retain more extended SFHs, although this model requires empirical rescaling of the metallicities due to known underestimation in the TNG suite \citep{Nelson2018}. Backsplash galaxies may naturally account for the presence of red quenched UDGs beyond the cluster virial radius \citep{Benavides2021}. The right panel of Figure \ref{fig:SFHs_all} shows the averaged SFHs of both quenched UDGs and star forming ones adapted from \citet{Benavides2024}. The authors compare the shape of these SFHs with the high-density and low-density ones from observations \citep{FerreMateu2023}, showing that quenched cluster UDGs in TNG50 have similar quenching timescales to observed high-density systems, whereas field and group UDGs display more prolonged star formation, consistent with observational trends. 

Across simulations, the environmental modulation of SFHs is also reflected in the MZR. Figure \ref{fig:mzr_sims} shows again the already discussed MZR (as in Figure \ref{fig:mzr_all}), this time including the values from the different suites of simulations just discussed. This figure, adapted from \citet{FerreMateu2023}, splits both observations and simulations by their local environment (high density for cluster, low density for field, groups or infall/filaments). We first note, as previously discussed, that some of the simulations like TNG first apply a shift in their total metallicities to match the observations. After those corrections, high-density simulations broadly follow the observed relation but show less scatter than observed UDGs, even if considering the entire simulated ensemble. Notably, no high-density simulation reproduces the lowest-metallicity systems interpreted as `failed galaxies'. In low-density environments, NIHAO simulations can achieve such low metallicities, although these systems cannot represent failed galaxies due to their extended, bursty SFHs. 

Overall, the comparison of SFHs and MZR across simulations and observations emphasises the combined influence of internal feedback, mergers, and environment in producing the full diversity of UDGs, while highlighting specific areas, such as extremely metal-poor UDGs, where simulations remain incomplete.

\subsection{Do UDGs build their stellar mass in the same way as classical dwarfs?}
Spatially resolved stellar populations offer critical insight into UDG formation, complementing integrated measurements. Observations of stellar population gradients, though limited, thus provide a more detailed view of these processes.

DF44 was the first UDG with measured radial gradients \citep{Villaume2022}, benefiting from high signal-to-noise and multiple radial bins. Its age profile is essentially flat, consistent with classical dwarfs, but the metallicity profile is flat-to-rising, contrasting with both classical dwarfs and simulations of UDGs. This is, DF44 seemed to have formed inside-out rather than outside-in like the rest of classical dwarfs. Subsequent spatially resolved studies confirmed that DF44 is not unique. Other cluster UDGs (DFX1, DF07, PUDG-R21, FCC\,224) and even NUDGEs (PUDG-R27 and VCC\,1448), all exhibit similarly flat age gradients and flat-to-rising metallicity profiles \citep{FerreMateu2025, Buzzo2025b, Levitskiy2025}, and the group UDG NGC1052-DF2 shows also comparable trends \citep{Fensch2019}. 

\begin{figure}
    \centering
    \includegraphics[width=0.99\linewidth]{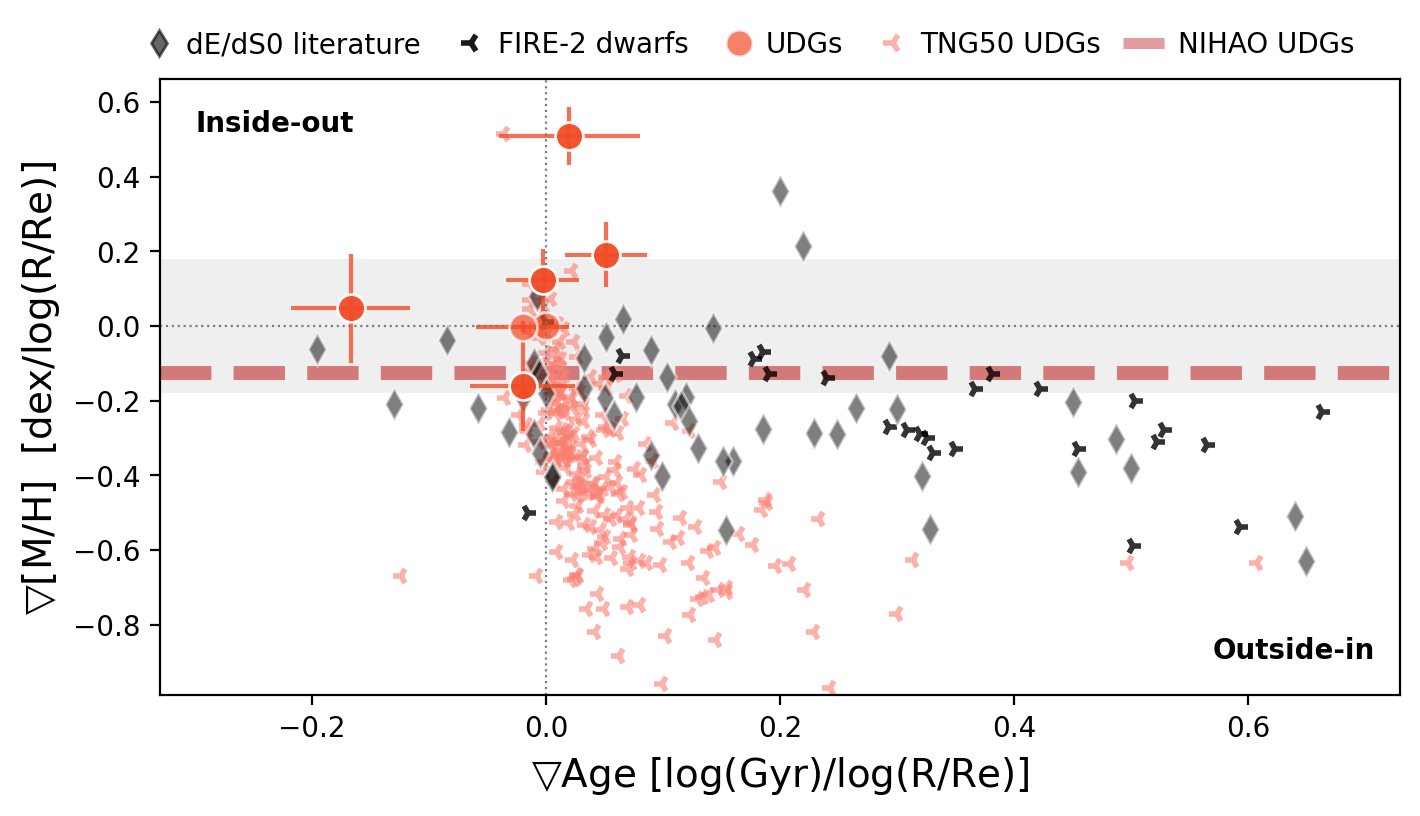}
    \caption{Figure adapted from \citet{FerreMateu2025} to show the age and metallicity gradients derived for UDGs in that work but also including the recent results of \citet{Buzzo2025b} and \citet{Levitskiy2025}. Individual simulated UDGs from TNG50 \citep{Benavides2024} and the mean gradient for NIHAO ones \citep{CardonaBarrero2022} are shown as red asterisks and a red dashed line, respectively, while FIRE-2 simulated dwarfs \citep{Graus2019, Mercado2021}, are shown in black asterisks. The grey band indicates the regime considered as a `flat' metallicity gradient according to \citet{Mercado2021}. Observed classical dwarfs are shown as black diamonds (data from \citealt{Koleva2011, Sybilska2017, Bidaran2022, Lipka2024a}). This figure shows that UDGs are broadly consistent with extending the distribution of both simulated and observed dwarfs, towards flat-to-rising metallicity profiles. This suggests an inside-out formation scenario, unlike most known dwarfs that are thought to form outside-in.
    }
    \label{fig:grads_sims}
\end{figure}

Figure\,\ref{fig:grads_sims} is updated from \citet{FerreMateu2025}\footnote{We caution the reader there was an error in the computation of the age gradients for the simulated TNG50 UDGs and FIRE-2 dwarfs in the original figure of \citet{FerreMateu2025} that we have amended here. We note, however, that the results and trends to not change, and thus no erratum is published.} 
to include all UDGs with spatially resolved stellar populations. It also summarises the gradients of simulated UDGs \citep{CardonaBarrero2022, Benavides2024} and classical dwarfs \citep{Graus2019, Mercado2021}, while presenting the observed distribution in classical dwarfs \citep{Koleva2011, Sybilska2017, Bidaran2022, Lipka2024a}. NIHAO UDGs display mild negative metallicity gradients, ($\bigtriangledown$([$Z$/H])$=-$0.125\,dex/log(R/R$\rm_{e}$), red dashed line), while TNG50 UDGs exhibit steeper gradients, particularly among star-forming systems (see \citealt{Benavides2024}). Notably, the only TNG50 UDGs that show slightly flat or rising metallicity profiles are those that have experienced tidal stripping.
Figure\,\ref{fig:grads_sims} illustrates that UDGs populate the tail of classical dwarf gradients, towards the flat-to-rising metallicity regime. Collectively, these patterns support a `\textit{Everything, everywhere, all at once}', nearly coeval, inside-out formation mode \citep{FerreMateu2025}. 

It is noteworthy that most of the UDGs studied for resolved gradients are cluster members and, in particular, GC rich, with the exception of NGC1052-DF2 in a group environment. Extending these studies to field UDGs and/or GC-poor systems will be essential to test whether these trends are universal or not. 

Finally, gradients in [Mg/Fe] remain largely unexplored. Currently, only DF44 has a published $\alpha$-element profile, showing a mildly declining gradient \citep{Villaume2022}. Incorporating [Mg/Fe] gradients across a broader sample would help clarify how UDGs assemble their $\alpha$ elements, and further distinguish what sets them apart from the general classical dwarf population.

\section{The Globular Cluster Systems of UDGs} \label{sec:GCs}

One of the most intriguing aspects of UDGs is their globular cluster (GC) content. Indeed, shortly after their initial discovery in the Coma cluster, they were found to host {\it ``extensive"} (i.e. rich) GC systems \citep{vanDokkum2017}. While many host GC systems similar in number to those of classical dwarfs (i.e. up to a dozen GCs), others reveal relatively rich GC systems (i.e. many tens, and perhaps up to a hundred). Thus, along with effective radius, UDGs have some properties that are more similar to giant galaxies, despite their dwarf-like stellar masses. As discussed below, the UDGs with rich GC systems may also have giant galaxy-like GC luminosity functions and dark matter halos. 

The largest samples of UDGs with GC counts, and the benefit of being at a similar distance (which reduces systematics), are those in the Coma (distance $\sim$ 100\,Mpc), Perseus (75\,Mpc), Hydra-I (50\,Mpc) and Fornax (20\,Mpc) clusters. Good quality GC studies also exist for UDGs in the Virgo cluster (16.5\,Mpc), although distance variations of a few Mpc ($\sim$20\%) can be present. Using studies of UDGs in these clusters and in lower density environments (e.g. the MATLAS survey \citealt{Marleau2021}), along with limited spectroscopy, we address here various questions pertaining to the GC systems of UDGS.

\subsection{Quantifying GC Systems}
Typically, the first step in measuring a GC system associated with a galaxy is to count its GCs from imaging. For UDGs beyond the Local Volume, ground-based imaging cannot resolve them (unless it is complemented with adaptive optics) and so it struggles to separate GCs from foreground stars and background galaxies even with multiple filters. \textit{HST}, \textit{JWST} and \textit{Euclid} have provided a significant improvement in selecting GCs by partially, or fully, resolving them  
(GCs typically have half-light radii of 2--3\,pc). \textit{Euclid} has 0.1" pixels in its VIS instrument and can partially resolve GCs at Virgo-like distances ($\sim$16.5 Mpc). On \textit{HST}, the ACS/WFC (0.05" pixels) and WFC3/UVIS (0.04" pixels) instruments can reach to $\sim$35 and 40\,Mpc respectively. The NIRCam instrument on \textit{JWST} in its highest resolution mode has a pixel scale of 0.03". Strictly speaking, imaging only gives GC candidates which require a radial velocity for confirmation. 

In order to estimate a total number of GCs (N$_{\rm GC}$), corrections are needed for magnitude and radial incompleteness, and background contamination. The magnitude correction may require an assumption about the GC luminosity function (GCLF), i.e. whether to use that of luminous galaxies or one more appropriate for dwarfs.  
Different studies have adopted different approaches and hence variations in the number of GCs assigned to a galaxy can vary from one author to the next. 

In order to compare different galaxies, it is useful to normalise N$_{\rm GC}$ by galaxy luminosity or stellar mass since bigger galaxies tend to have larger GC systems. Galaxy luminosity has the advantage of being an observable (when the distance is known), but it is difficult to compare young galaxies with older, passive hosts. In principle, using galaxy stellar mass (M$_{\ast}$) removes the effects of stellar populations on the luminosity. This is combined with the total mass of the GC system M$_{\rm GC}$ to create the ratio M$_{\rm GC}$/M$_{\ast}$. This ratio, or a similar variant, is becoming more widely used than the older specific frequency (S$_N$) which is essentially the ratio of N$_{\rm GC}$ to galaxy luminosity L$_V$, i.e. S$_N$ = N$_{\rm GC}$ $\times$ 10$^{0.4(M_V + 15)}$.

To calculate M$_{\rm GC}$ it is common to simply multiply N$_{\rm GC}$ by the mean mass of a GC. This has the advantage of being less sensitive to the GCs with the faintest magnitudes.
A mean GC mass of 2 $\times$ 10$^5$ M$_{\odot}$ \citep{2007ApJS..171..101J}
is often assumed, 
although second order variations in the `universal' GCLF are known \citep{2010ApJ...717..603V}. For example, although the GCLF in dwarf galaxies typically appears to have a Gaussian shape \citep{2009MNRAS.396.1075G}, some studies suggest that the mean GC mass is closer to 1 $\times$ 10$^5$ $M_{\odot}$ for dwarf galaxies. 
Therefore given the low numbers of GCs in the lowest mass dwarf galaxies, it is better to sum individual GC masses rather than assume a mean value  \citep{2018MNRAS.481.5592F}. 

\begin{figure}
    \centering
    \includegraphics[width=0.8\linewidth, angle=-90]{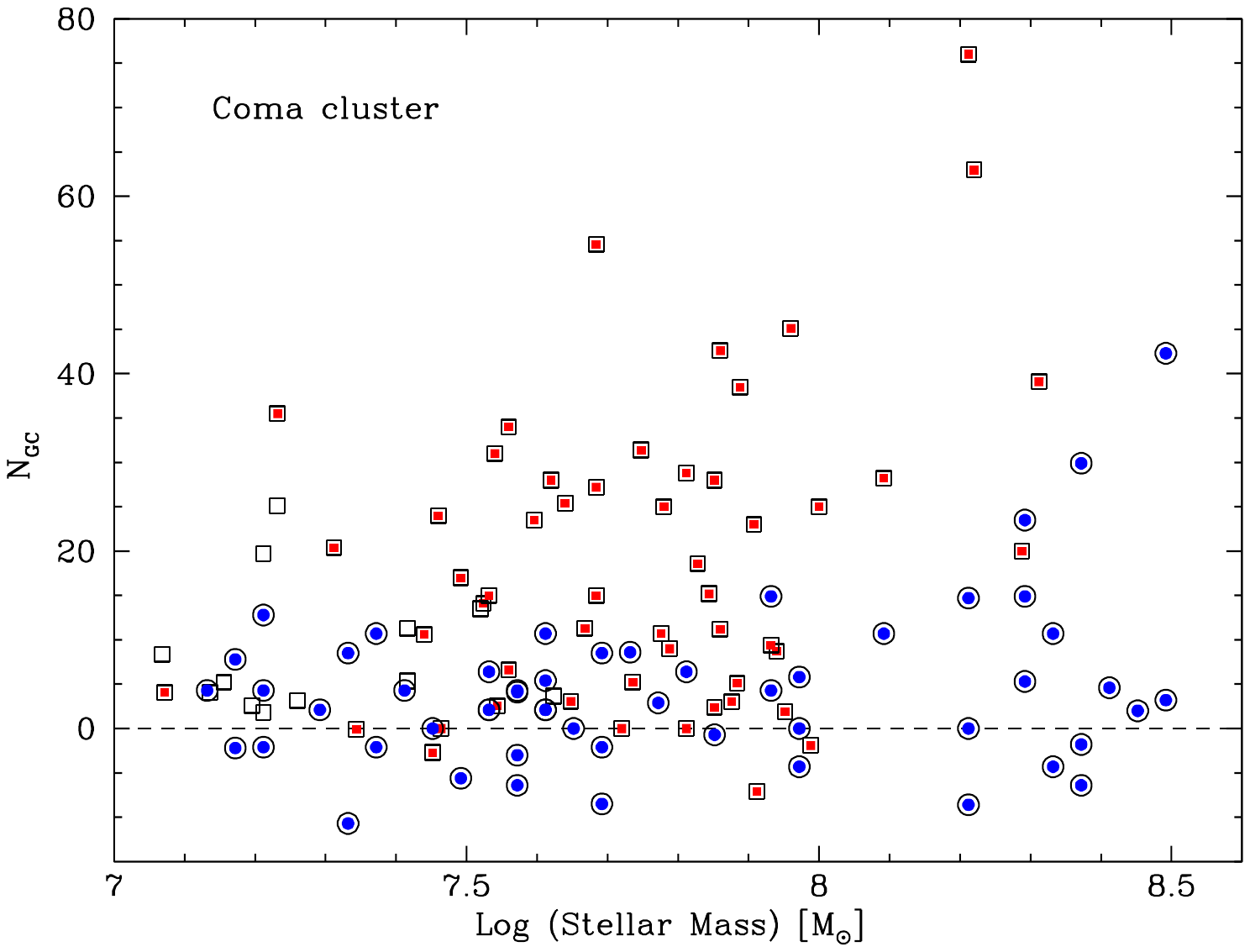}
    \caption{Number of GCs around Coma cluster galaxies as a function of galaxy stellar mass. Red squares show UDGs, blue circles classical dwarf galaxies and open squares non-UDG low surface brightness galaxies. There is a clear trend for Coma UDGs to have more GCs, on average, than a Coma classical dwarf galaxy of the same stellar mass. Data from \cite{Forbes2020}.
    }
    \label{fig:gcnum}
\end{figure}

One of the key differences between UDGs and classical dwarfs is that in some cases, such as in the Coma cluster, UDGs can host a larger number of GCs, on average, for a given host galaxy stellar mass. 
One of the first works to indicate this was \cite{Beasley2016b}, who found a rich GC system in the Coma UDG DF17 from 3-filter HST imaging.
Figure~\ref{fig:gcnum} shows GC counts for Coma cluster galaxies with data from \cite{Forbes2020}. Here all of the GC counts have been estimated using \textit{HST} imaging in a similar manner and the galaxies are located at the same distance (i.e. 100\,Mpc in the Coma cluster), so it is a good dataset for a relative comparison between UDGs and classical dwarfs. Although GC-poor UDGs overlap with classical dwarfs, the GC-rich ones extend to numbers not seen in dwarfs.  The addition of further data has indicated that the most GC-rich Coma UDGs are {\it not} simply puffed-up classical dwarf galaxies. 

While the Coma cluster reveals the best evidence for GC-rich UDGs, there are strong indications for an environmental dependency,  with GC richness increasing with environmental density. In the field and groups, UDGs generally host small numbers of GCs (\citealt{2020ApJ...902...45S};
\citealt{Jones2023}). In the MATLAS survey of low density groups, around 2/3 of their  74 UDGs imaged hosted between zero and two GCs, and only four had more than 20 GCs. 
The Fornax cluster with a halo mass of $\sim$7 $\times$ 10$^{13}$ M$_{\odot}$ is about 1/10 the mass of the Coma cluster (and contains around 1/10 as many UDGs). In a study of UDGs (and LSBs) in the Fornax cluster, \cite{Prole2019} found all of the dozen UDGs to have fewer than 20 GCs. With a comparable halo mass to the Coma cluster, the Perseus cluster and has been studied using both \textit{HST} \citep{Janssens2024} and \textit{Euclid} \citep{Saifollahi2025}. In both studies several UDGs hosting up to 40--50 GCs were reported. And similar to Coma, the UDGs in Perseus have more GCs per unit stellar mass than classical dwarfs \citep{Saifollahi2025}. More work is needed, along with comparative studies of classical dwarfs, to quantify these environmental trends.  

The trend of higher GC richness for UDGs in higher density environments is similar to that found for classical dwarfs, e.g. \cite{2008ApJ...681..197P}. For classical dwarfs, this may be explained by different assembly histories for cluster and field galaxies, combined with quenching due to cluster infall 
\citep{2016MNRAS.455.2323M}. 
The expectations for GC richness in the various UDG formation mechanisms are summarised in Figure~\ref{fig:glance_table}.  For example, the UDG formation processes involving high spin or SN feedback of classical dwarfs would be expected to leave the number of GCs relatively unchanged. Tidal heating \citep{Carleton2021} may boost the M$_{\rm GC}$/M$_{\ast}$ ratio of UDGs by the preferential loss of stars but not the absolute number. On the other hand, while mergers may add more GCs, the ratio M$_{\rm GC}$/M$_{\ast}$ in the merger remnant will be similar to that of the progenitors since additional stars are added as well. Tidal stripping of a classical dwarf must be severe to remove GCs but certainly will not raise their absolute number. Elevating the GC count compared to classical dwarf galaxies might be possible in both the `bullet dwarf' and `failed galaxy' scenarios. We note that in the UDG literature, both a simple GC number count (e.g. N$_{\rm GC}$ $>$ 20) and M$_{\rm GC}$/M$_{\ast}$ (e.g. $>$ 2\%) have been used as a measure of GC richness.

\subsection{Do UDGs have a `universal' Gaussian GCLF?}
{ The GCLF turnover magnitude occurs at slightly fainter magnitudes \citep{2010ApJ...717..603V} in dwarf compared to giant galaxies. This is likely 
driven by differences in the ages and metallicities of the GCs in dwarf galaxies. } The effect is much weaker in blue filters (e.g. the $g$-band) than red filters (e.g. $z$-band).
For example, from a sample of luminous ellipticals the Vega system turnover magnitude in the $I$-band is M$_I\sim -8.5$ \citep{2001AJ....121.2950K} while it is M$_I\sim -8.2$ for a sample of dwarf ellipticals \citep{2007ApJ...670.1074M}. A similar GCLF offset in magnitude was noted by \cite{2025A&A...697A..10S} after stacking massive vs. dwarf galaxies in the Fornax cluster  (although they also noted three dwarf galaxies to have brighter than normal GCLF turnovers). 

Studies of the GCLF of UDGs are hindered by typically low numbers of GCs for individual galaxies and by distance, which restricts both the magnitude coverage of the GCLF and the ability to resolve individual objects with space-based telescopes. Using deep \textit{HST}/ACS imaging, \cite{2025ApJ...978...93M} studied the rich GC system of the UDG VCC\,615 in the Virgo cluster (N$_{\rm GC}$ = 25). This galaxy has the advantage of a Tip of the Red Giant Branch measured distance and resolved GCs with 92\% coverage of the GCLF magnitude range. They found a turnover magnitude of M$_I$ between $-$8.45 and $-$8.68 for their sample of GCs, thus closer to that of luminous ellipticals than dwarfs.

\begin{figure}
    \centering
    \includegraphics[width=0.95\linewidth]{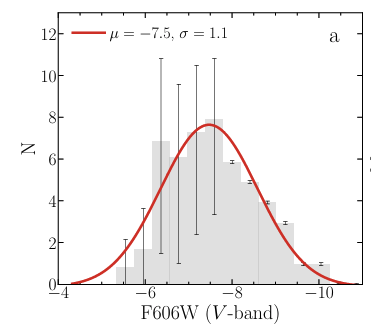}
    \caption{GCLF of NGC\,5846\_UDG1 reproduced from figure 4 \textit{left} of \protect{\citet{Danieli2022}}. This UDG has a GCLF that is well fit by a GCLF with a turnover magnitude of M$_V=-7.5$ and a dispersion $\sigma=1.1$ (i.e. similar to that of giant galaxies). Thirty GC candidates are spatially resolved by \textit{HST}/WFC3 with sizes consistent with GCs at a distance of 26.5\,Mpc.}
    \label{fig:udg1}
\end{figure}

Perhaps the best defined GCLF of an individual UDG is that of NGC\,5846\_UDG1. It is located in the NGC\,5846 group and its GC system has been probed by \cite{Danieli2022} using deep \textit{HST}/WFC3 imaging. They identified GC candidates covering 94\% of the GCLF and estimated a total system of 54$\pm$9 GCs. From an initial dozen confirmed GCs with radial velocities \citep{Muller2020}, this has risen recently to 20 \citep{2025MNRAS.539..674H}. Furthermore, some 30 candidates having measured FWHM sizes consistent with GCs at 26.5 Mpc (the assumed distance used by \citealt{Danieli2022}).   
At this assumed distance the measured turnover corresponds to M$_V=-7.5$ (see Figure~\ref{fig:udg1}), i.e. consistent with that of luminous galaxies \citep{2012Ap&SS.341..195R}. Using a stricter colour selection for GC candidates, 
\cite{2025arXiv251209990G} argue for a smaller system of 
33 $\pm$ 3 GCs and a closer distance of 20 Mpc. At this distance, UDG1 would lie outside of the NGC 5846 group, thus making UDG1 even more remarkable -- a GC-rich field UDG.

In a study of 18 UDGs (and other LSB galaxies) in the Perseus cluster, \cite{2025ApJ...984..147L} found that the majority have a GCLF turnover consistent with that of dwarf galaxies. However, they also found two galaxies to have brighter GCLF turnovers by $\sim$1\,mag. A similar $\sim$1\,mag brighter than the typical dwarf galaxy GCLF was reported by \cite{Janssens2022} for the field UDG DGSAT-I. By stacking the GCLF of six Coma cluster UDGs observed using \textit{HST}/ACS, \cite{Saifollahi2022} measured a turnover that was more consistent giant galaxy GCLFs (once a correction for 
Vega to AB magnitudes is correctly applied). 
In the case of the GC-rich UDG (N$_{GC}$ = 52$\pm$12 studied  
by \cite{2023ApJ...954L..39F}, 
the GC candidates are more compatible with a dwarf galaxy GCLF. 

Thus there appears to be a diversity of GCLFs in UDGs, with some revealing dwarf-like turnover magnitudes and others being more consistent with more massive galaxies. As well as these single Gaussian GCLFs (albeit of varying turnover magnitudes), there are several reported cases of double-peaked GCLFs. In particular, the brighter peak corresponds to GCs with luminosities similar to those of UCDs (i.e. up to 2-3 magnitudes brighter than the brightest GCs). Such objects have been identified around the galaxies DF2 and DF4 in the NGC\,1052 group, along with a population of normal luminosity GCs \citep{vanDokkum2019}. The status of the overly luminous GCs, and their large in physical sizes, depends on the distance to the host galaxy which has been disputed by some (e.g. \citealt{Trujillo2019}). However, when considered along with the other fainter GC candidates in these two galaxies their GCLFs are double-peaked, and will remain so regardless of their assumed distance \citep{2021ApJ...909..179S}. 

A galaxy with many similar properties to DF2 and DF4, including the presence of some overly-luminous GCs is FCC224, i.e \cite{Tang2025b}. In this case, the galaxy lies in the outer parts of the Fornax cluster within its velocity phase-space envelope and with an surface brightness distance distance, thus it can be confidently assigned to the Fornax cluster. \cite{Tang2025b} showed the GCLF of FCC224 along with that of DF2 and DF4, in comparison to the stacked GCLF of classical dwarf galaxies in the Fornax cluster. The GCs of FCC224 reveal a $\sim$1\,mag brighter GCLF turnover (with several of the brighter GCs spectroscopically-confirmed by \citealt{Buzzo2025b}) and a few fainter GCs, with the GCLF of the classical Fornax dwarfs lying at intermediate magnitudes. 

Another key property connecting DF2 and DF4 to FCC224 is the lack of dark matter in the galaxy inner regions. While distance affects the calculation of enclosed dynamical mass via the galaxy effective radius, it does not affect the velocity dispersion which for all galaxies is remarkably low at 5--10\,km\,s$^{-1}$. Whether the presence of overly-luminous GCs is associated with a DM-free UDG or not is yet to be fully tested and requires further examples. For the two UDGs with well-defined GCLFs mentioned above,  NGC5846\_UDG1 (see Figure \ref{fig:udg1}) and 
VCC615, there is no evidence for a population of overly-luminous GCs. 

In conclusion, for some UDGs the GCLF turnover magnitude is more similar to dwarf galaxies (i.e. galaxies that match UDGs in stellar mass). However, for others, the GCLF is in closer agreement to more massive galaxies (which might suggest a match in total halo mass). There appear to be strong examples of both dwarf-like and giant-like GCLFs for UDGs. Thus UDGs exhibit a diversity in their GCLF properties, including the presence of some overly-luminous GCs and double-peaked GCLFs.

\renewcommand{\arraystretch}{1.5}%
\begin{table}
    \centering
	\caption{Summary of UDG Globular Cluster System Radial Extent}
	\label{tbl:rgc}
	\begin{tabular}{ccc} % four columns, alignment for each
		\hline
	Sample	 & R$_{\rm GC}$/R$_{\rm e}$ & Ref.\\
		\hline
        VCC615 & 1.47$^{+0.58}_{-0.36}$ &   \cite{2025ApJ...978...93M}\\
        NGC5846\_UDG1 & 0.8$\pm$0.2 &  \cite{Danieli2022}\\
        Field/Group & $\sim$1 &  \cite{Marleau2024}\\
        Coma & $\sim$1.5 & \cite{vanDokkum2017b}\\
        Coma & $\sim$1.5 & \cite{Lim2018}\\
        Coma & 1.09$\pm$0.14 & \cite{Saifollahi2022}\\
        Perseus & $\sim$1.2 & \cite{Janssens2024}\\
        Perseus & $\sim$0.72 & 
        \cite{Saifollahi2025}\\
Fornax & $\sim$1.7 &  \cite{Prole2019}\\       
        \hline
	\end{tabular}
\end{table}

\subsection{What is the radial extent of GC systems?}

Another key property of GC systems is their radial extent (usually quantified as the half-number radius) compared to that of the galaxy stars (effective radius), i.e. R$_{\rm GC}$/R$_{\rm e}$. This ratio often plays a crucial role in determining the correction for radial incompleteness and hence the total GC system counts. As the fraction of accreted GCs is reduced in lower mass galaxies, the ratio R$_{\rm GC}$/R$_{\rm e}$ also gets smaller 
(\citealt{2017MNRAS.472L.104F}, \citealt{2024ApJ...966..168L}). 
A key question is whether this ratio is different to those of classical dwarf galaxies. In a recent study of a large sample of classical dwarf galaxies, \cite{Carlsten2022} found R$_{\rm GC}$/R$_{\rm e}$ to vary from 1.06, on average, for Local Volume dwarfs to 1.25 for Virgo cluster dwarfs. 

From deep \textit{HST} imaging of the individual UDGs, VCC\,615 and NGC\,5846\_UDG1, 
\cite{2025ApJ...978...93M} and \cite{Danieli2022} reported R$_{\rm GC}$/R$_{\rm e}=1.47^{+0.58}_{-0.36}$ and 0.8$\pm$0.2, respectively. For UDGs in the low density environments of the MATLAS survey, \cite{Marleau2024} found a median value of R$_{\rm GC}$/R$_{\rm e}\sim1$ from \textit{HST}/ACS imaging. For UDGs in the Coma cluster, \cite{Lim2018} using \textit{HST}/ACS concluded that R$_{\rm GC}$/R$_{\rm e}$ was consistent with 1.5. A similar conclusion was reached by \cite{vanDokkum2017b} for a small sample Coma UDGs. However, using that same data, \cite{Saifollahi2022} derived 
R$_{\rm GC}$/R$_{\rm e}=1.09\pm0.14$. 
Different approaches conducted by individual researchers have led to different results for UDGs in the Perseus cluster as well. In an \textit{HST} study of the Perseus cluster, \cite{Janssens2024} found a range of 1 to 2.5 in R$_{\rm GC}$/R$_{\rm e}$, with an average value of $\sim$1.2 for their UDGs. Exploiting \textit{Euclid}'s wide field-of-view, 
\cite{Saifollahi2025}
found an average for their four stellar mass bins to be R$_{\rm GC}$/R$_{\rm e}\sim0.72$. In a study of Fornax cluster UDGs \cite{Prole2019} found a median of  R$_{\rm GC}$/R$_{\rm e}\sim1.7$. 

\cite{Saifollahi2025} 
were also able to determine relative differences between samples of UDGs and non-UDG galaxies at a constant stellar mass, with good number statistics.  
For the stellar mass range $6.5<$log(M$_{\ast}$/M$_{\odot}$)$<8$, for four mass bins they found R$_{\rm GC}$/R$_{\rm e}$ to vary from 0.52 to 1.13 for UDGs and from 0.73 to 1.20 for non-UDGs. The simple average of the four bins was 0.72 and 1.04 for UDGs and non-UDGs, respectively. Thus, in a relative sense, the GC profiles of UDGs appear to be slightly more compact than those of non-UDGs of the same stellar mass. This argues against UDGs simply being puffed-up dwarfs \citep{Saifollahi2022}. 
Compared to classical dwarfs, both the galaxy effective radius (by definition) and the GC system half-number radius of UDGs are relatively large. 

In summary, classical dwarf galaxies appear to have GC systems that are slightly more extended, on average, than their galaxy stars (i.e. R$_{\rm GC}$/R$_{\rm e}\ge 1$). Whereas the GC systems of UDGs (see Table 2) tend to be more compact, with R$_{\rm GC}$/R$_{\rm e}\le 1$. For both UDGs and classical dwarfs, a variety of GC profiles exist.  

\subsection{What Properties do S$_N$ and M$_{\rm GC}$/M$_{\ast}$ Correlate With?}

Specific frequency (S$_N$) and the ratio of GC system mass to galaxy stellar mass (M$_{\rm GC}$/M$_{\ast}$) are proxies for the efficiency of both GC formation and destruction. So it is interesting to understand what host galaxy properties they might correlate with.   

%As noted in the Introduction, UDGs are bias against elongated galaxies (low b/a ratios). 
\cite{2024MNRAS.529.4914P} collected GC and host galaxy axial ratios for UDGs in the Virgo, Coma and Perseus clusters along with field/group environments from the MATLAS survey. They showed that high S$_N$ (GC-rich) UDGs tend to be rounder on average, with almost a complete lack of GC-rich UDGs with b/a $\le$ 0.6. Given this trend was also seen in the MATLAS sample of field/group UDGs \citep{Marleau2024} it is unlikely to be purely an environmental process. This trend was interpreted by \cite{2024MNRAS.529.4914P} as a reflection of relatively inefficient GC formation in rotationally-supported UDGs. More measurements of the rotation in UDGs, over a range of GC richness, should allow this prediction to be tested. 

Using \textit{HST}/ACS, \cite{Lim2018} examined the dependence of S$_N$ on various properties for Coma UDGs. They found weak trends for S$_N$ to be higher in the cluster centre and in smaller sized UDGs. There was no clear trend with galaxy surface brightness in their data. \cite{Forbes2020} combined the data of \cite{vanDokkum2017}, \cite{Lim2018} and  
\cite{Amorisco2018} to create an enlarged sample of 85 UDGs in Coma with GC counts. They found weak trends for higher S$_N$ in smaller sized, fainter surface brightness and bluer UDGs.  
We note that strong trends exist between galaxy luminosity and properties such as effective radius and colour, so trends with S$_N$ (effectively scaling inversely with luminosity) should be treated with caution. In a recent study of UDGs and classical dwarfs in the Perseus cluster, \cite{2025arXiv251211070T}
did not find any strong GC trends with position in the cluster for their 40 UDGs. However, they did find that lower surface brightness and larger UDGs tend to be more GC-rich.  

\begin{figure}
    \centering
    \includegraphics[width=1.05\linewidth]{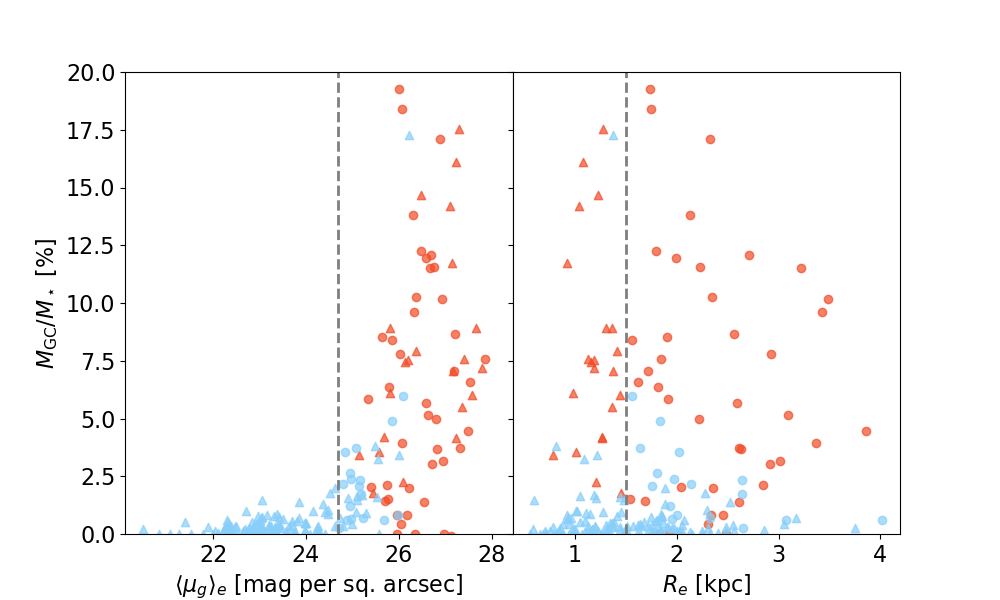}
    \caption{ M$_{\rm GC}$/M$_{\ast}$ vs average $g$-band surface brightness within the effective radius ($<\mu_g>_e$) and effective radius R$_{\rm e}$. The plot shows UDGs (circles) and non-UDGs (triangles) in both Coma (blue) and Perseus (red) clusters. Vertical lines show the UDG definition. Data are from \cite{Forbes2020} for Coma and \citet{Tang2026} for Perseus. The Coma cluster reveals higher ratios than the Perseus cluster galaxies. The trend with R$_{\rm e}$ is rather weak but a strong trend for higher ratios in lower surface brightness galaxies is evident. 
    }
    \label{fig:mgc}
\end{figure}

In Figure~\ref{fig:mgc} we show M$_{\rm GC}$/M$_{\ast}$ vs surface brightness and effective radius for UDGs and non-UDGs in the Coma and Perseus clusters from the above studies. While there is a strong trend of higher ratios in lower surface brightness (lower stellar density) UDGs, there is only a very weak trend for the highest values in the smallest sized UDGs. Additionally, \cite{Marleau2024} studied the S$_N$ trends for UDGs in the MATLAS survey using \textit{HST}/ACS. While they found no significant trend with colour, surface brightness or axial ratio, they did find higher S$_N$ in larger-sized UDGs (i.e. opposite to the trends for Coma cluster UDGs mentioned above). 

In terms of stellar populations, higher M$_{\rm GC}$/M$_{\ast}$ ratios are associated with slightly older and more metal-poor UDGs \citep{Forbes2025}. This is consistent with them forming efficiently at earlier times. However, the sample sizes are currently restricted to around a dozen UDGs with both spectroscopy and GC counts. More data are required to further test the initial findings that GC-rich UDGs have stellar populations resembling those of metal-poor GCs, as might be expected if the stars of such galaxies are dominated by disrupted GCs.  
Ideally this data should be spectroscopic but the colours of GCs and their host galaxy can be a useful proxy for their stellar populations. 

\subsection{Do UDGs follow the GC number\,--\,halo mass Relation?} \label{sec:gcnum_halomass}

It is well-known that normal galaxies follow the relation between GC number and halo mass, with a scatter of $\sim$0.3 dex, over many orders of magnitude, e.g. \cite{2009MNRAS.392L...1S}, \cite{2015ApJ...806...36H}, 
\cite{Harris2017}, 
\cite{2018MNRAS.481.5592F}, \cite{2025ApJ...978...33L} and \cite{Zaritsky2022b}. 
{We note that a similar near-linear relation exists between the mass of the nuclear star cluster in UDGs and the halo mass \citep{Khim2024}. 
Such a relation offers an alternative method for estimating the halo mass of UDGs if they host a nuclear star cluster. It also suggests a close relationship between nuclear star clusters and the GC systems of UDGs.}

{Although simulations exist that can reproduce the observed GC relation for normal galaxies,  e.g. \cite{2014ApJ...796...10L}, \cite{2017MNRAS.472.3120B}, 
\cite{Burkert2020} and \cite{2021MNRAS.505.5815V}, they have yet to fully explore the situation for UDGs. }
Using Illustris TNG100, \cite{Carleton2021} examined the relation between GC system mass and halo mass for the UDGs in that simulation suite. They found that they systematically deviate from the observed relation and with larger scatter. 
We also note the work of \cite{Doppel2024} in which they tagged dark matter particles as GCs within the Illustris TNG50 simulation. However, as all of their galaxies, including UDGs, have fewer than 20 GCs and lie within dwarf-like dark matter halos, the regime of GC-rich UDGs was not explored. 

\begin{figure}
    \centering
    \includegraphics[width=0.98\linewidth]{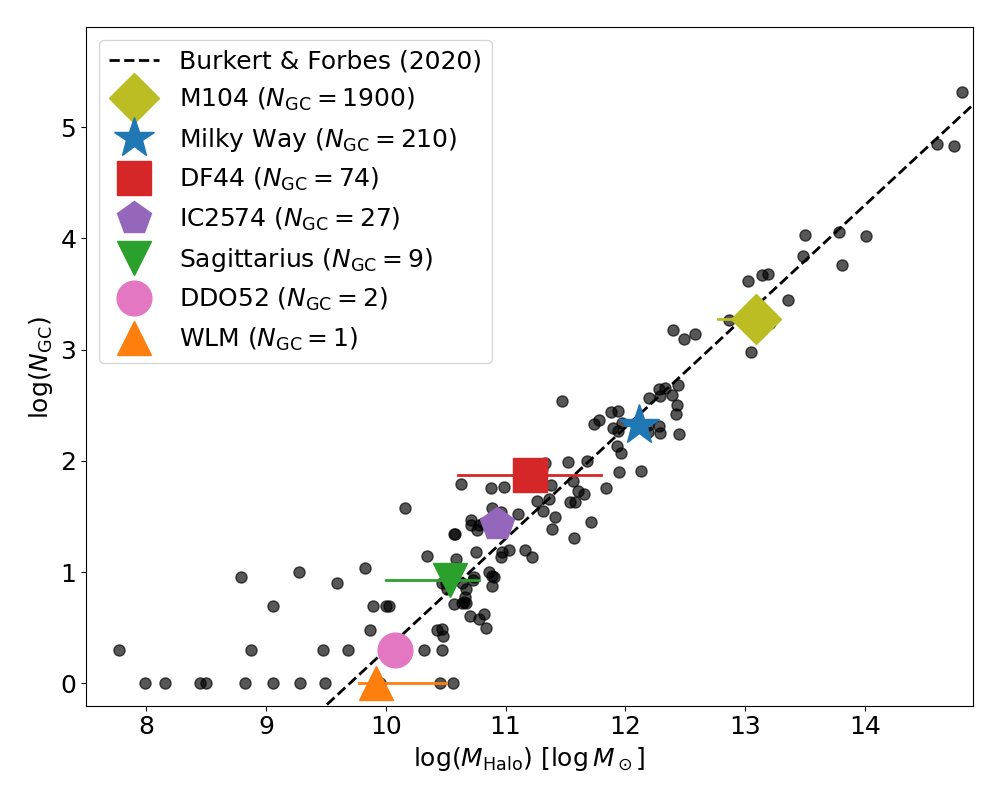}
    \caption{Scaling relation between the number of globular clusters and the halo mass of a galaxy reproduced from \citet[][their figure 1]{Forbes2025b}. Black symbols show normal galaxies with a range of morphology and environment taken from \protect{\citet{Burkert2020}}. The dashed line shows their linear fit of 5 $\times$ 10$^9$ M$_{\odot}$ in halo mass per GC. Coloured symbols show three UDGs and two NUDGes with independent halo mass estimates. The  Milky Way and Sombrero galaxies are  shown for comparison. GC uncertainties are smaller than the symbol size. The UDGs follow the relation of normal galaxies.
    }
    \label{fig:ngcmh}
\end{figure}

Other than the well-studied UDG DF44 with an independent halo mass measurement \citep{vanDokkum2019b}, it is unknown whether other UDGs also follow the GC number\,--\,halo mass relation. Recently, \cite{2025MNRAS.543L...1F} revisited this issue using independent halo masses from the literature for an additional three nearby UDGs and two NUDGes (which have UDG-like sizes but slightly brighter than the UDG definition). They have GC counts ranging from a single GC associated with the WLM galaxy to the 74 of DF44. In this work it was found that UDGs do indeed follow the standard GC number\,--\,halo mass relation, as reproduced in Figure \ref{fig:ngcmh}. 

An interesting corollary of UDGs following the GC number\,--\,halo mass relation is that they reveal a systematic deviation in the stellar mass\,--\,halo mass relation with increasing GC count. The GC-poor UDGs are still consistent with the latter scaling relation (as might be expected if they are simply puffed-up dwarf galaxies) however the GC-rich ones are not. As found by \cite{Forbes2024}, the GC-rich UDGs have more massive halos and/or fewer stars than expected if they were to follow the standard stellar mass\,--\,halo mass relation. Such galaxies that have failed to form as many stars as expected for their halo mass have been dubbed failed galaxies (as described Section 3.3.4).  
%(\citealt{Danieli2022}, \citealt{Forbes2024}). 

\subsection{When did the first GCs form?\\}

Age measurements of GCs in UDGs are few and far between. The one GC associated with the WLM dwarf is very old, with an age comparable to the age of the Universe \citep{1999ApJ...521..577H}. From a high quality spectrum, \cite{2014A&A...565A..98L} found it to be consistent with an age of $\sim$13\,Gyr and very metal-poor ([Fe/H]$=-1.97$\,dex), which matches the age\,--\,metallicity relation of the WLM galaxy stars \citep{2013ApJ...767..131L}. The star formation history of WLM reveals a slow enrichment over cosmic time, resulting in a mean metallicity for the galaxy some 0.7\,dex higher than for its GC. In the case of UDG1 in the NGC~5846 group, several GCs have reported ages in \cite{2020A&A...640A.106M}, with a mean GC age of 11.2$^{+1.8}_{-0.8}$\,Gyr (which was similar to that of the galaxy field stars within the estimated uncertainties). 

Future work should include efforts to obtain ages and metallicities of GCs hosted by UDGs. The first epoch of GC formation appears to be relatively uniform irrespective of the host galaxy 
(e.g. see review by \citealt{2018RSPSA.47470616F}). However, determining the absolute ages, and hence when the first GCs formed, has been problematic even for 
Milky Way GCs. Although GCs were known to be very old, it was not entirely clear if they formed during, or after,  re-ionisation ($z\sim6$). With the advent of \textit{HST} 
(e.g. \citealt{2017MNRAS.467.4304V}), and more recently \textit{JWST} observations of lensed high redshift galaxies, a population of bright, compact star clusters was revealed. Their structural properties, ages, and metallicities all suggest that they are the high redshift counterparts of today's GCs, e.g. \cite{2023MNRAS.520L..58F}. Furthermore, the stars in some high redshift galaxies have revealed abundance patterns that resemble local GCs 
\citep{2025arXiv250511263N}. 

A summary of these high redshift observations is given in \cite{2024arXiv240521054A} and \cite{2025MNRAS.536.1878P}. Notable is the Sparkler at $z=1.378$ \citep{2022ApJ...937L..35M}, and two systems at redshifts $z>6$, i.e. Firefly at $z=8.304$ \citep{2024Natur.636..332M} and Cosmic Gems at $z=9.625$ 
\citep{2025arXiv250718705M}.
The mean age derived for the oldest star clusters are 4.20$\pm$0.26\,Gyr for the 5 GCs in the Sparkler, 102.4$\pm$7.9\,Myr for the 10 Firefly clusters, and 13$\pm$2.4\,Myr for the 5 Cosmic Gems clusters (we note that this is a rapidly moving field with new analysis and new data that may revise these values, e.g. 
\citealt{2025A&A...699A.240T}).
Using Ned Wright's Cosmology Calculator 
(https://www.astro.ucla.edu/wright/CosmoCalc.html) the look-back time corresponding to the redshift of each galaxy for the `General' Universe is an age of 13.721\,Gyr, i.e. 9.086\,Gyr for the Sparkler, 13.106\,Gyr for the Firefly, and 13.217\,Gyr for Cosmic Gems. The GC formation epoch is therefore 13.286$\pm$0.260\,Gyr, 13.208$\pm$0.008\,Gyr and 13.230$\pm$0.002\,Gyr for the Sparkler, Firefly, and Cosmic Gems respectively. 

\begin{figure}
%   \centering
    \includegraphics[width=0.49\textwidth]{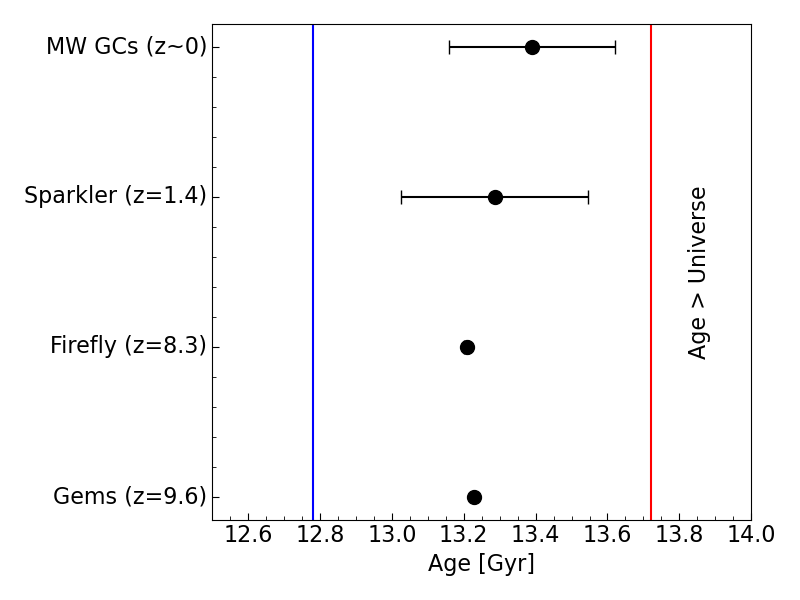}
    \caption{The first epoch of globular cluster formation. The red line shows the age of the Universe (13.72\,Gyr), and the blue line shows the age corresponding to z = 6 (near the end of re-ionisation). Symbols show the oldest GCs in 3 lensed systems and for a compilation of Milky Way GCs (with their systematic uncertainty). Globular clusters first formed $\sim$400 Myr after the Big Bang and before the end of re-ionisation. 
    }
    \label{fig:ages}
\end{figure}

Improvements in fitting and modelling the CMDs of Milky Way GCs have brought uncertainties down from 1--2\,Gyr to a fraction of that for sample of GCs. One example is the recent analysis by\cite{2025arXiv250319481V} of 68 Milky Way GCs with \textit{HST} photometry to derive a mean age for the oldest, metal-poor GCs of 13.39\,Gyr. The error budget is dominated by systematics of $\pm$0.23\,Gyr, while the statistical uncertainty is only $\pm$0.10\,Gyr. A summary of the ages of the oldest GCs is given in Figure~\ref{fig:ages}. As such it shows that the oldest GCs formed some 400\,Myr after the Big Bang but almost a Gyr before the end of re-ionisation at redshift $z\sim6$. There is no current evidence to suggest that the oldest GCs associated with UDGs are any different. 
%Nevertheless efforts should be made to obtain spectra, of sufficiently high S/N, to obtain the ages of the GCs around UDGs.  

The formation epoch of the first GCs has particular relevance for the `failed galaxy' scenario. If the GCs in UDGs form early as described in Figure~\ref{fig:ages}, make a significant contribution to the galaxy's halo 
\citep{Peng2016} via disruption, and the UDG evolves passively there after, then we would expect the field stars of a failed galaxy UDG to reveal GC-like stellar populations, i.e low metallicity and very old ages.

\subsection{How efficiently are GCs formed and destroyed?}

As mentioned, the number (or mass) of GCs per galaxy luminosity (or stellar mass) today is a combined measure of GC formation and destruction efficiency. 
The efficiency of producing gravitationally bound star clusters has been tackled in simulations \citep{2012MNRAS.426.3008K}, including high redshift galaxies \citep{2015MNRAS.454.1658K}. Assuming that star formation is an extension of that observed locally, but to higher gas fractions and pressures, then more efficient GC formation can be expected. This can be quantified as the fraction of the total luminosity of each galaxy contributed by the star clusters. As summarised by \cite{2025MNRAS.536.1878P}, this ratio can reach $\sim$75\% for some high redshift galaxies as observed by \textit{JWST}. In other words, up to three quarters of the luminosity produced in these high redshift galaxies comes from their system of star clusters. 

This luminosity ratio is related to the initial M$_{\rm GC}$/M$_{\ast}$ ratio via rather uncertain mass-to-light ratios, but nevertheless one can expect mass ratios to be of a similar order. \textit{JWST} has thus confirmed that GCs can form very efficiently at early epochs (a few Myr after the Big Bang). Further evidence of star cluster dominated star formation at early times comes from \textit{JWST} observations of super-solar N/O abundances, which are similar to those measured in Milky Way GCs \citep{2025arXiv250512505J}. Such vigorous star cluster formation would be expected to have an influence on the galaxy's evolution in the form of feedback of energy, and perhaps expulsion of the gas, which may now quench subsequent star, and cluster, formation. This quenching of star formation by star clusters is one possible mechanism to explain failed galaxy UDGs \citep{Danieli2022}, since the galaxy `fails' to form additional stars.  

Over cosmic time a large fraction of the initial GC population is expected to be removed by either internal (relaxation or mass loss) or external (tidal shock) processes. As well as reducing the number of GCs, these destruction processes 
{appear to operate in a similar manner preferentially destroying low mass clusters at early times, resulting in a final GCLF with the near-universal log-normal shape and turnover magnitude \citep{2018RSPSA.47470616F}. However, detailed modelling of the GCLF over time and the contribution of disrupted GCs to the host galaxy’s stars remains to be investigated.} As the GCs are disrupted, their stars contribute to the main body of the galaxy. This disruption of GCs over cosmic time has been modelled in classical dwarf galaxies by \cite{2024MNRAS.527.2765M}. They found lower (or inefficient) GC disruption rates in lower mass, and lower density, galaxies. Thus, for a given initial GC formation efficiency, we might expect higher M$_{\rm GC}$/M$_{\ast}$ ratios in lower mass dwarf galaxies and in lower surface brightness galaxies such as UDGs (see also \citealt{Forbes2025}). 

In summary, the elevated GC counts seen in UDGs may be a combination of efficient formation at early times and relatively inefficient destruction. 
However, detailed modelling of the GCLF over time and the contribution of disrupted GCs to the host galaxy’s stars is still required.

\subsection{Do UDGs have the same stellar populations as their GC systems?}

The concept of UDGs being `pure stellar halos' of old metal-poor stars was originally proposed by 
\cite{Peng2016}. This may arise since GCs are disrupted over cosmic time, with their stars contributing to the field stars of the host galaxy. If few, if any, in-situ field stars formed, then the stellar mass of a galaxy could be dominated by GC-like stellar populations. Some support for this idea comes from the spectroscopic observations of NGC5846\_UDG1 by \cite{2020A&A...640A.106M}. Using MUSE they found that the host galaxy stars and a dozen GCs shared the same mean old age and low metallicity. A similar situation was reported by 
\cite{Buzzo2025b} for the stars and half dozen GCs of FCC224.  
In the case of NGC1052-DF2 however, \cite{Fensch2019} found the galaxy stars to have a higher metallicity than the co-added spectra of GCs. Spectroscopy of cluster UDGs found some to have GC-like stellar populations, although the majority tend to be slightly younger and more metal-rich, on average, than typical  GCs \citep{FerreMateu2023}.

Deriving the stellar populations of individual GCs is time consuming and therefore samples are extremely limited. An alternative approach is to directly compare the relative colours of UDGs and their GCs. For old ages, colour is largely an indicator of metallicity but its sensitivity depends on the filter combination.  In an \textit{HST} study of six UDGs, \cite{Saifollahi2022} found GCs and their host UDG to have the same colour on average (some being bluer, some redder). For a sample of a dozen UDGs in the Perseus cluster with \textit{HST} imaging, \cite{Janssens2024} found a mean colour offset of 0.07$\pm$0.08, i.e. the host galaxy was redder than the GCs on average but there was no significant difference when the uncertainty was taken into account. A redder galaxy colour would suggest it is more metal-rich (e.g. enriched by additional star formation) than its metal-poor GCs. There was no obvious trend between the colour difference and the total number of GCs in each UDG. \cite{vanDokkum2020} studied the colours of the GCs in both DF2 and DF4 in the NGC\,1052 group using \textit{HST}. They found the luminous GCs to have monochromatic colours indicating very similar ages and metallicities for the GCs in both galaxies. The host galaxies had slightly redder colours consistent with the higher metallicity inferred from spectroscopy \citep{Fensch2019}. For the field UDG DGSAT-I, \cite{Janssens2022} found identical colours for the GCs and galaxy stars. 

In summary, the field stars of some UDGs have the same stellar populations to their GCs (i.e. very old and metal-poor). However, in other cases the galaxy stars have an elevated metallicity (and perhaps slightly younger ages) suggesting some in-situ star formation in addition to disrupted GC stars. Future work should include more studies to measure the relative colour difference of field stars and the GC system for individual UDGs. This difference can then be correlated with UDG properties, e.g. the ratio of mass in GCs today to the stellar mass of the UDG. Failed galaxy UDGs might be expected to reveal GC-like stellar populations, whereas UDGs that are puffed-up dwarf galaxies may reveal more enriched and younger stars.  

\subsection{Does Cluster Infall Affect GC Richness?}

Early infall of galaxies into a cluster is expected to quench subsequent star formation. Thus the earlier the infall, the higher is the expected fraction of bound mass in old GCs relative to the galaxy's field stars  \citep{2016MNRAS.455.2323M}. Early infall, combined with early and efficient GC formation, in the TNG100 simulation of 
\cite{Carleton2021} are the main reasons that model UDGs have twice as many GCs as non-UDGs. 
There is indeed some observational evidence for this effect in classical dwarf galaxies
\citep{2008ApJ...681..197P}, 
however for UDGs, the evidence is less convincing. \cite{Lim2018} examined how specific frequency of GCs (S$_N$) varied with projected cluster-centric radius for Coma UDGs. For one luminosity range they found a weak trend of higher S$_N$ at small projected radii but no clear trend for their other two luminosity ranges.

In a heterogeneous study of UDGs in the Hydra~I, Perseus, Coma and Virgo clusters, \cite{Forbes2023} studied their location in the infall diagram 
\citep{2017ApJ...843..128R} 
as a function of GC richness with a division at N$_{\rm GC}\ge20$. The expected trend was not present. A similar lack of a trend was found by \citet{Tang2026} in their study of 40 UDGs in the Perseus cluster. Using M$_{\rm GC}$/M$_{\ast}$ as a measure of GC richness, they found a weak tendency for GC-rich UDGs to be found in later infall regions, contrary to the expectation that they occupy early infall regions. 
Furthermore, the quenching timescale of GC-rich UDGs does not seem well aligned with their infall time (\citealt{FerreMateu2023, Doll2025}) 
These results might suggest some other process is responsible for GC richness (e.g. quenching prior to cluster infall).

\section{Are UDGs different from Classical Dwarf Galaxies?}\label{sec:twoclasses}
At this point, it is reasonable to ask whether UDGs are simply classical dwarf galaxies with lower surface brightness and larger sizes on average, or if they represent something fundamentally different. In many respects, UDGs do resemble classical dwarfs—for instance, their global stellar population properties are comparable (Section \ref{sec:stellar_pops}). Yet, important differences emerge. One striking example is the on-sky distribution of their elongation: while dwarfs show a fairly uniform b/a distribution, UDGs are biased toward larger b/a ratios.  If UDGs are inherently disky in morphology, this may be due to a bias towards face-on orientations, a bias that stems from their definition as low-surface-brightness systems (see further \citealp{2024MNRAS.529.4914P}). {Alternatively, UDGs may not be inherently disky in morphology}. Moreover, although their average metallicities are similar to those of classical dwarfs, some UDGs deviate significantly from the canonical MZR. Their metallicity gradients also tend to be flat or even rising, in contrast to the falling gradients usually observed in dwarfs. Furthermore, some UDGs host far more GCs—both in absolute terms and relative to stellar mass than typical dwarfs. This overabundance implies overly massive halos, inconsistent with a simple dwarf-galaxy nature. In this regard, at least a fraction of UDGs clearly diverge from the classical dwarf population.

As surveys have expanded and follow-up spectroscopic and SED-fitting studies followed suite, the diversity among UDGs became increasingly evident in their structure, stellar populations, and GC properties. With ongoing and upcoming all-sky surveys such as \textit{Euclid}, and \textit{LSST}, the number of detected UDGs is expected to grow into the tens of thousands. Indeed, \textit{Euclid} has already transformed our view of low-mass galaxies: early results indicate that about 8\% of dwarfs are new UDG candidates, corresponding to roughly 200 new systems (see e.g. \citealt{Marleau2025a,Marleau2025b,Saifollahi2025}). This rapid increase makes it essential to develop methods for distinguishing between different UDG types.

\begin{figure}
    \centering
    \includegraphics[width=0.9\linewidth]{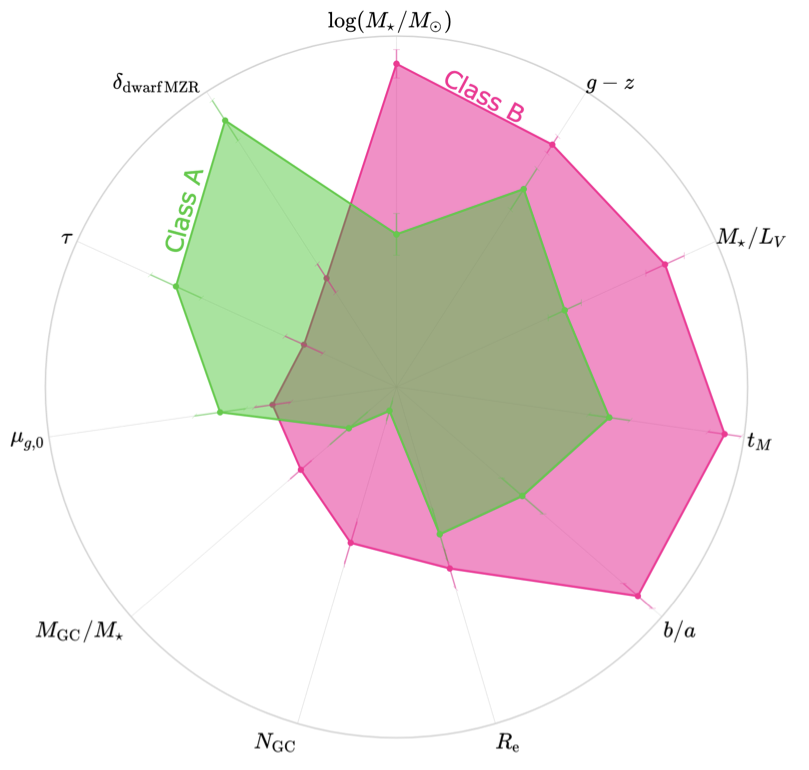}
    \caption{Figure reproduced from figure 5 of \citet{Buzzo2025}, which shows the result of performing the clustering algorithm of {\tt KMeans} using the stellar populations and internal properties of UDGs. The figure shows how two distinctive types of UDGs are found. Given the overall properties of each group, group A seems to be more compatible with the puffed dwarf formation channel, while the class B would be more representative of `failed galaxies' one.}
    \label{fig:classes}
\end{figure}

One promising approach is the application of machine learning techniques, as demonstrated by \citet{Buzzo2025}. Using a sample of 88 UDGs in both field and cluster environments, they applied the {\tt KMeans} clustering algorithm to classify UDGs according to their structural and stellar population properties. Figure \ref{fig:classes} reproduces their figure 5, presenting these properties in a polar plot. This figure reveals two clearly distinct classes of UDGs. Class A UDGs are the least massive, with lower surface brightness, elongated shapes, more compact sizes, and very poor GC systems. They are typically bluer, with younger stellar populations, lower stellar mass-to-light ratios, more extended star formation histories, and metallicities that follow the classical dwarf MZR. These features suggest that Class A corresponds to the `puffy dwarf' formation channel. In contrast, Class B UDGs are more massive and of higher surface brightness, with rounder shapes, larger sizes, and rich GC systems, yielding high GC-to-stellar mass ratios. They are generally redder, with older stellar populations, higher mass-to-light ratios, early-quenched star formation histories, and metallicities that fall below the classical dwarf MZR. Such galaxies have been proposed to represent the `failed galaxy' formation pathway.

A small caveat to this result is that the MATLAS survey has since revised its surface brightnesses (\citealp{Marleau2021}; Marleau, et al. Erratum Submitted, \citealt{Poulain2025b}). \citet{Buzzo2025}'s classification of galaxies followed that of the MATLAS survey as published in \citet{Marleau2021}, and so those galaxies considered to be a `UDG' by MATLAS, and hence by \citet{Buzzo2025}, were based on inaccurate surface brightnesses. (It does not affect the KMeans analysis that used the \citet{Buzzo2025} surface brightness measurements, not those from the MATLAS survey; it merely affects the classification of a galaxy as a UDG or not). In this revision, all UDGs previously called so, remain so, however 14 new galaxies now fit the UDG regime.  

We note that this distinction holds when including NUDGe objects, although it starts to dilute the trends due to a wider spread in the age and metallicities. Nonetheless, as discussed to this point and shown by Figure\,\ref{fig:classes}, the most influential features in separating between the two classes seem to be the deviation from the classical dwarf MZR, the number of GCs, and the galaxy shape (given by the axis ratio b/a). It will be interesting to keep updating this classification as time evolves and more UDGs are observed and followed up, for instance, adding the [Mg/Fe] ratio in the classification, which is currently lacking in many of the studies.

\section{Some Future Directions for UDG Research} \label{sec:future}
\subsection{UDGs and their Relationship to Environment}

\cite{Janssens2017} discovered that UDGs and UCDs had an inverse spatial relationship in the A2744 cluster, suggesting that the disruption of nucleated UDGs left UCD remnants and contributed field stars to the intra-cluster light (ICL).  Support for this process came from \cite{Wang2023} who followed the evolution of nucleated UDGs in the Virgo cluster. Thus evidence is accumulating that disrupted UDGs could be a contributor to the ICL (and the UCD population). From measurements of individual stars in the ICL of the  Virgo cluster, \cite{2007ApJ...656..756W} found the ICL to have a range of metallicities but it was dominated by stars of low metallicity ([$Z$/H]$<-1$\,dex) and old age, similar to the stellar populations of cluster UDGs, e.g. \citealp{FerreMateu2023}). Further study of the relationship between UDGs, UCDs and the ICL is thus warranted to best understand the interplay between these different objects.

Furthermore, currently the studies of UDG number with environmental mass (e.g., Figure \ref{fig:nudg_m200}) combine different studies that use a mix of definitions as to what a UDG is. These studies also make different corrections to account for the incompleteness of imaging to the cluster's virial radius. Clearly, this is inappropriate for accurately determining the slope of the relationship between UDG number and environmental mass. A consistent approach to studying this relationship will aid efforts to understand the rate at which UDGs are created/destroyed by tidal forces. Studying the evolution of the number of UDGs in clusters with redshift will also help to understand how tidal forces may form/destroy UDGs, as well as their accretion rates into clusters. Current studies of UDGs in clusters beyond $\sim100$Mpc are limited to \citet{Janssens2017, Janssens2019, Carleton2023, Ikeda2023}. With a plethora of deep \textit{HST} and \textit{JWST} exposures of clusters at a wide range of redshifts, along with new images expected from \textit{Euclid}, it should be possible to expand these efforts. 

\subsection{UDG's Resolved Internal Properties}

Recently, measurements of stellar population radial gradients in UDGs have become available in the literature (see Section \ref{sec:stellar_pops}). Such gradients tend to be relatively flat in both age and metallicity. This contrasts with the negative metallicity gradients usually seen in classical dwarfs and predicted for UDGs by simulations. A constant old age and low metallicity might however, be expected if disrupted GCs dominate the stellar populations of the host galaxy. More work is needed to expand the samples of GC-poor vs. GC-rich comparisons, and to search for other secondary parameters that might affect stellar population gradients in UDGs, such as rotation, size, and environment. Gradients of the $\alpha$ elements are largely unconstrained observationally at this time but hold important clues regarding timescales of star formation. 

Just as data regarding stellar population gradients in UDGs is sparse, so too are radial kinematic profiles. Until recently, DF44 and NGC~1052-DF2 are the only 2 UDGs with radial velocity dispersion profiles measured \citep{vanDokkum2019b, Emsellem2019}. Rotation has now been detected in several UDGs (e.g., \citealp{Chilingarian2019, Buttitta2025, Levitskiy2025}), which intriguingly is frequently found to be prolate. Samples are currently small, and the radial extent probed is also limited; it is likely that peak rotation has not yet been reached. This will also allow the effects of galaxy inclination to be explored and, hence, determine how many of the UDGs with no detected rotation are truly non-rotating systems. 

Furthermore, resolved kinematic profiles can be used to infer resolved mass profiles for UDGs and can be used to fit dark matter halos. In this regard, UDGs may be perfect for constraining alternative theories of dark matter. UDGs are sufficiently extended to measure the dark matter profile out to larger radii than classical dwarfs. They are also frequently found to be extremely dark matter dominated; as such, much of their kinematics is dominated by interactions within their dark, not baryonic, matter. Thus, any interactions such as the soliton expected in fuzzy dark matter (e.g., \citealp{Wasserman2019}) or dark matter composed of ultra-light axions (e.g., \citealp{Montes2024}), may be constrained by studying UDGs. Likewise, UDGs may provide constraints on non-cold dark matter, alternative gravity theories (e.g., \citealp{Muller2019}). More work is needed in this realm.

\subsection{Advancing the Formation Theories of GC-rich UDGs}
The GC systems of UDGs have been the subject of much study, yet key questions remain. Understanding why they host up to several times more GCs than their classical dwarf galaxy counterparts may shed crucial insight into how UDGs form and evolve. Most current simulations predict standard dwarf-like GC systems, i.e. they do not predict the GC-rich UDGs nor overly massive halos (a consequence of the GC number--halo mass relation). Thus new simulations are needed that produce overly massive halo (or equivalently stellar-suppressed) UDGs. One possible approach is the `genetic' modification of initial conditions as carried out by \cite{2019ApJ...886L...3R} for very low mass galaxies. Here quenching and late assembly suppresses star formation resulting in overly massive halo galaxies relative to the standard stellar mass--halo mass relation. Could this `failed galaxy' process be responsible for the GC-rich UDGs? And, if so, what other properties are predicted for UDGs formed in this way?

The role of cluster infall on GC systems also needs to be explored further. Although GC system trends within a cluster appear weak, there are hints of cluster-to-cluster variations. Namely, GC-rich UDGs appear to be more frequent in the largest clusters like Coma, than the smallest, such as Fornax. Will these results be verified with systematic wide-field studies of GC systems out to the virial radius in several clusters?

We also see a diversity in GC system properties from their radial extent to their luminosity functions. What drives this diversity? And what is the origin of the overly-luminous GCs seen in some UDGs?

\subsection{Going Darker: Identifying Even More Extreme Galaxies}
A promising new approach to identify even lower surface brightness galaxies is to use their (relatively high surface brightness) globular clusters as `lamp posts'. This approach was demonstrated by \cite{Li2022} who used a spatial statistical technique to identify an overdensity of GCs with no associated host galaxy light in the \textit{HST} imaging of the  Perseus cluster. Subsequent imaging from \textit{Euclid} revealed some diffuse emission at the location of the GCs, and hence a very faint host galaxy. They estimated the ratio of the light in the GC system to that of the galaxy to be a remarkable 16--33\%. For galaxies located closer than the Perseus cluster, at $\sim$75 Mpc, it may be possible to obtain radial velocities for the GCs. This would confirm their physical status and may allow for a system velocity dispersion to be measured (leading to an enclosed dynamical mass). 

While several telescopes offer wide areas that facilitate the search for additional UDGs (e.g. \textit{VST, Euclid, Rubin}), an alternative method is the detailed follow-up from a blind HI survey. Radio telescopes such as Arecibo, Parkes, ASKAP, and recently FAST, have conducted HI surveys which have, initially at least, detected a number of extragalactic HI sources that lack an obvious optical counterpart. {Primarily, follow-up deep imaging is likely to find low surface brightness galaxies that are star forming at some low level, thus being different to those studied in this review. However, an interesting, outstanding question for modern galaxy formation is when such gas reservoirs fail to form a galaxy, or form a galaxy that is largely quiescent for the age of the Universe and thus is ``dark''.} Recently, Monaci et al. (submitted) used the public FAST catalogue to generate a list of $\sim$60 nearby dark galaxy candidates. Deeper optical imaging may reveal a very low surface brightness UDG or a place tight upper limit on the stellar mass of the galaxy. Dark galaxies with no stars but significant HI gas are expected in large numbers in the $\Lambda$CDM framework.  

\subsection{Other Wavelengths and New Facilities}

The study of UDGs with other techniques and/or wavelengths beyond the optical/IR and radio has been limited to date. Even if stacking of many UDGs may only provide an upper limit, e.g. weak lensing \citep{Sifon2018} or X-rays \citep{Kovacs2019}, further efforts are encouraged. If UDGs reside in massive dark matter halos, they may be expected to exhibit an excess {of ionised gas emitting photos in the X-ray/far UV} in comparison to smaller halos. Likewise, if UDGs' field stars are significantly composed of stars from GCs, there may be an expectation of an excess of stars in X-ray binaries in comparison to normal dwarfs. Both scenarios require deep high energy exposures to test. 

It is worth noting that there are a large number of deep-wide surveys that are beginning, or are soon to begin, surveying the sky (e.g., \textit{Euclid} and \textit{LSST}). These surveys should present an order of magnitude more UDGs across large swathes of the sky, and build a far more complete picture of the UDG population.

Sadly, however, the future prospects are more challenging for expanding UDG spectroscopic efforts. Thirty-meter class telescopes will be needed to build a well-sampled, statistical sample of UDG spectroscopic properties to truly understand their formation. Until these telescopes are completed, innovative imaging methods on the available deep/wide data (e.g., the SED fitting) and machine learning classification approaches like KMeans are the only way to build and study statistical UDG samples. Spectroscopy will likely need to be limited to detailed studies of the most interesting cases to understand the extreme, rather than the mean, of UDGs.

\section{Conclusions} \label{sec:conclusions}
In this Dawes Review, we have summarised the last decade of research into ultra-diffuse galaxies (UDGs). We have placed particular emphasis on those UDGs that are quiescent, as the initial discovery of UDGs in the Coma cluster was focused on UDGs such as these. We have summarised the general properties of UDGs (Section \ref{sec:general_properties}), the proposed formation scenarios for UDGs (Section \ref{sec:formation_scenarios}), UDG internal kinematic properties (Section \ref{sec:internal_properties}), UDG internal stellar population properties (Section \ref{sec:stellar_pops}) and the globular cluster (GC) systems of UDGs (Section \ref{sec:GCs}). A brief discussion as to the main differences between UDGs and classical dwarf galaxies as well as the possible future of the field, was included in Sections \ref{sec:twoclasses} and \ref{sec:future}. Below we list the main items of note for each section of our review:

In terms of the general properties of UDGs we highlight:

\begin{itemize}
    \item The original UDG definition is {based on somewhat arbitrary criteria and selects galaxies that are large and faint. Compared to the handful of galaxies found prior to 2015, a surprisingly large number of galaxies have been found to fit these criteria}. {The selection is biased towards} galaxies that are redder and rounder ({perhaps} viewed more face-on). {Less-strict} definitions for UDGs that have been used may create a bias to select galaxies forming via more varied formation pathways. 

    \item UDGs are found in all environments. Given the relationship between UDG numbers and environmental mass is approximately linear, the environment likely does not play a strong net role in the creation/destruction of UDGs. In particular, UDGs' frequent lack of tidal features, even once stacked, provides little evidence for recent tidal interactions. 
    
    \item UDGs currently comprise $\sim5-8\%$~of all galaxies. Their inclusion is necessary to accurately reproduce galaxy mass/luminosity functions. 
\end{itemize}

In terms of the formation scenarios for UDGs we highlight:

\begin{itemize}
    \item We have identified at least 11 scenarios that have been proposed to form a UDG and provided a visual summary of each (Figure \ref{fig:glance_table}). From the discussion of internal properties in Sections \ref{sec:internal_properties}, \ref{sec:stellar_pops} and \ref{sec:GCs}, it is obvious that no single scenario can explain the entire population, with many UDGs clearly forming from distinct scenarios to one another. Further, while many of these scenarios are consistent with UDGs being simply the large size end of the normal dwarf galaxy distribution (i.e., puffy dwarfs) others require different dark matter halo characteristics. These latter galaxies may be thought of as dark matter halos that have failed to form their expected stellar mass (i.e., failed galaxies) and may host 
    large numbers of GCs.
\end{itemize}

In terms of the internal kinematic properties of UDGs we highlight:

\begin{itemize}
    \item Current UDG dynamical masses provide preliminary evidence that UDGs do not follow stellar mass\,--\,halo mass relationships established prior to their discovery. They indicate that UDGs likely follow the GC number\,--\,halo mass relationship (see also below and Figure \ref{fig:ngcmh}). A corollary of this is that many UDGs likely reside in cored dark matter halos. 

    \item When placed on the fundamental plane of elliptical galaxies UDGs reside in their own region of parameter space. This is primarily driven by their large half-light radii.
\end{itemize}

In terms of the internal stellar population properties of UDGs we highlight:

\begin{itemize}
    \item UDGs span a wide mean age range (see Table \ref{tab:1}), but with metallicities generally below solar ([$Z$/H]$\sim–1.1$\,dex). When placed in the canonical mass\,--\,metallicity (MZR) relation, the bulk of UDGs scatter around the local dwarf galaxy MZR (`puffy dwarfs'), while a minority of very metal--poor UDGs follow instead the $z\sim2$ MZR (`failed galaxies'). Rare examples with the most elevated metallicities are compatible with being tidally stripped UDGs.  

    \item  Overall, UDGs show above solar $\alpha$-enhancements ([Mg/Fe]$\sim0.4$\,dex). However, some extreme outliers show [Mg/Fe]$>1$\,dex — unprecedented in any galaxy type, raising questions about enrichment/depletion mechanisms.
    
     \item Local environment seems to play a role on how UDGs build up their stellar mass: UDG in higher density environments are typically older, show short and early quenched SFHs, and have the most elevated [Mg/Fe]. On the contrary, UDGs in lower density environments tend to have extended SFHs, later quenching, lower metallicities and [Mg/Fe].

     \item In the few spatially resolved cases, (GC-rich) UDGs in clusters have relatively flat age profiles, but surprisingly flat-to-rising metallicity gradients. This is unlike classical dwarfs and simulated UDGs. It implies a rather co-eval, inside-out growth and assembly pathways distinct from classical dwarfs.

\end{itemize}

In terms of the GC systems of UDGs we highlight:

\begin{itemize}
    \item UDGs have been shown in many environments (e.g., the Perseus/Coma clusters) to have, on average, more GCs than dwarf galaxies of a similar stellar mass. Alternatively, this presents as UDGs having a higher $M_{\rm GC}/M_\star$~ratio than other galaxies.

    \item UDGs have been shown to challenge the conventional wisdom of a `universal' GCLF. While some UDGs indeed follow the dwarf-like GCLF, other UDGs have been shown to have a more giant-galaxy-like GCLF. Furthermore, some UDGs have been shown to have over-luminous GC systems, and/or have a GCLF that is best fit by a dual-peaked Gaussian. 

    \item UDGs appear to have GC systems of slightly smaller relative radial extent (i.e., $R_{\rm GC}/R_{\rm e}$) than dwarf galaxies of a similar stellar mass. While this trend appears relevant to the UDG population as a whole, individual UDGs may deviate with many shown to have radially extended GC systems. 

    \item Current evidence for the handful of UDGs with independent halo mass measurements favours their following the GC number\,--\,halo mass relationship. A corollary of this finding is that many GC-rich UDGs cannot follow the stellar mass\,--\,halo mass relationship due to their following the GC number\,--\,halo mass relationship. 
    
\end{itemize}

We conclude our review by thanking all those researchers whose hard work, dedication and creativity have led to the past decade of progress in understanding ultra-diffuse galaxies. We eagerly anticipate a second decade of the mystery, discovery and camaraderie that is provided by the study of UDGs.

%%%%%%%%%%%%%%%%%%%%%%%%%%%%%%%%%%%%%%%%%%%%%%%%%%

\section*{Acknowledgements}

We thank the dual anonymous referees for their comments, which helped us to best frame this exciting field of research. We thank the AGATE team members A. Romanowsky, J. Brodie, Y. Tang, A. Levitskiy, M. Monaci, D. Vaz and H. Christie for their support and insight in the creation of this review. We thank Y. Tang for providing his data to include in our Figures. We thank the ARC for financial support via DP220101863 and DP250101673. A.F.M acknowledges support from RYC2021-031099-I and projects PID2021-123313NA-I00 and PID2024-162088NB-I00 of MICIN/AEI/10.13039/501100011033/FEDER,UE, NextGenerationEU/PRT. 

%%%%%%%%%%%%%%%%%%%%%%%%%%%%%%%%%%%%%%%%%%%%%%%%%%
\section*{Data Availability}
This article is a review of work that is publicly available. In particular, this work makes use of a publicly available catalogue of UDG spectroscopic properties available \href{https://github.com/gannonjs/Published_Data/tree/main/UDG_Spectroscopic_Data}{here}. The full references for the catalogue are \citet{mcconnachie2012, vanDokkum2015, Beasley2016, Martin2016, Yagi2016, MartinezDelgado2016, vanDokkum2016, vanDokkum2017, Karachentsev2017, vanDokkum2018, Toloba2018, Gu2018, Lim2018, RuizLara2018, Alabi2018, FerreMateu2018, Forbes2018, MartinNavarro2019, Chilingarian2019, Fensch2019, Danieli2019, vanDokkum2019b, torrealba2019, Iodice2020, Collins2020, Muller2020, Gannon2020, Lim2020, Muller2021, Forbes2021, Shen2021, Ji2021, Huang2021, Gannon2021, Gannon2022, Mihos2022, Danieli2022, Villaume2022, Webb2022, Saifollahi2022, Janssens2022, Gannon2023, FerreMateu2023, Toloba2023, Iodice2023, Shen2023, Janssens2024, Gannon2024, Buttitta2025, FerreMateu2025, Levitskiy2025, Buzzo2025b}.

\bibliographystyle{mnras}
\bibliography{bibliography}

\end{document}